\documentclass[oneside,11pt]{article}

\usepackage[english]{babel}    
\usepackage{tgpagella}
\usepackage{enumerate} 
\usepackage{float}
\usepackage{graphics,latexsym,amsfonts}       
\usepackage{amssymb,amsthm,hyperref,mathtools}   
\usepackage[dvipsnames]{xcolor} 
\usepackage{mathrsfs} 
\usepackage{graphicx} 
\usepackage{float} 
\usepackage{ulem}    
\usepackage{picture}
\hypersetup{
    colorlinks, 
    linkcolor={red!50!black},  
    citecolor={blue!50!black},  
    urlcolor={blue!80!black}   
}
\usepackage[longnamesfirst]{natbib}    
\usepackage{tabularx}

\def\reference#1{\href{#1}{Cliquer ici pour voir une r\'ef\'erence.}} 
\usepackage{sectsty} 

\usepackage{setspace}
\usepackage{caption}
\sectionfont{\normalfont\scshape\centering}
\subsectionfont{\centering}
\providecommand{\U}[1]{\protect \rule{.1in}{.1in}}
{\normalfont\itshape}

\usepackage[top=1.25in,bottom=1.25in,left=1.25in]{geometry}
\def\reference#1{\href{#1}{Click to see a reference}} 

{\color{Maroon}}

\def\X{\mathscr{X}}
\def\es{\varnothing}

\def\sbs{\subseteq}

\DeclareMathOperator*{\argmax}{argmax} 
\makeatletter

\newtheoremstyle{mytheoremstyle} 
    {\topsep}                    
    {\topsep}                    
    {\itshape}                   
    {}                           
    {\scshape}                   
    {.}                          
    {.5em}                       
    {}  

\theoremstyle{mytheoremstyle}
\newtheorem{theorem}{Theorem} 
\newtheorem*{theorem*}{Theorem}

\newtheorem{lemma}{Lemma}
\newtheorem{corollary}{Corollary} 
\newtheorem*{corollary*}{Corollary} 
\newtheoremstyle{mydefinitionstyle} 
    {\topsep}                    
    {\topsep}                    
    {}                   
    {}                           
    {\scshape}                   
    {.}                          
    {.5em}                       
    {}  
\theoremstyle{mydefinitionstyle}
\newtheorem{definition}{Definition} 

\newtheorem*{question*}{Question}

\newtheorem{remark}{Remark}

\usepackage{pstricks}
\usepackage{pst-grad} 
\usepackage{pst-plot}
\usepackage{pst-node}
\usepackage{pst-text}

\makeatletter

\title{Limited attention and models of choice:\\ A behavioral equivalence\thanks{
 We are grateful to the Editor and the Associate Editor of this journal who handled the manuscript, and to the two anonymous referees.
Their advice has helped us make significant improvements to the paper.
We are indebted to Jean-Paul Doignon, Paolo Ghirardato, Alfio Giarlotta, M.\,Ali Khan, Marco Mariotti, and Daniele Pennesi for several comments and suggestions.
We also thank the audiences of the 2025 Decision: Theory, Experiments, and Applications Workshop at the Institute Henri Poincar\'{e} of Paris, the XLVII Annual Meeting of the Italian Association for Mathematics Applied to
 Social and Economic Sciences at the University of Milano Bicocca, and the 2024 Workshop on Economic Theory at Bocconi University.
Angelo Petralia acknowledges the support of "Ministero del Ministero dell'Istruzione, dell'Universit\`a e della Ricerca (MIUR), PE9 GRINS "Spoke 8", project \textit{Growing, Resilient, INclusive, and Sustainable}, CUP E63C22002120006.} }

\author{ 
\textsc{Davide Carpentiere}\thanks{University of Catania, Catania, Italy. davide.carpentiere@phd.unict.it.}, \textsc{Angelo Enrico Petralia}\thanks{University of Catania, Catania, Italy. angelo.petralia@unict.it.}   
\thanks{Corresponding author.}}

\usepackage{fancyhdr} 
\date{}
\begin{document} 
\maketitle

\begin{abstract}
\noindent 
 We show that many models of choice can be alternatively represented as special cases of \textit{choice with limited attention} \citep{MasatliogluNakajimaOzbay2012}, singling out the properties of the unobserved attention filters that explain the observed choices.
For each specification, information about the DM's consideration sets and preference is inferred from violations of the contraction consistency axiom, and it is compared with the welfare indications obtained from equivalent models.
Remarkably, limited attention always supports the elicitation of DM's taste arising from alternative methods.
Finally, we examine the intersections between subclasses, and we verify that each of them is independent of the others.
\end{abstract}

\medskip


\medskip


\section*{Introduction}\label{SECT:Introduction}

 Limited attention in individual choice has been extensively analyzed in marketing, economics, and psychology (see \citealt{ChiangChibNarasimhan1999}, \citealt{GilbrideAllenby2004}, and, more recently, \citealt{SmithKrajbic2018} for some references).
According to this paradigm, the decision maker (DM) does not examine all available options.
Instead, she picks the preferred item from her \textit{consideration set}, i.e. the alternatives to which she pays attention.
This selection bias has been originally described by \cite{MasatliogluNakajimaOzbay2012}, who justify a choice function by means of a linear order, and an \textit{attention filter}, i.e. the collection of subsets, each one associated to some menu, on which the DM maximizes her preference.\footnote{Later on, \cite{Cattaneoetal2020} proposed a generalization of limited attention that recovers stochastic choice data.}
Attention filters are interpreted as the DM's consideration sets, and they satisfy a unique property, which requires that the removal from a menu of an unconsidered item does not alter the DM's attention.
Despite its elegance, limited attention can hardly be tested in experiments with few alternatives.
Indeed, this pattern justifies any choice on three items, and more than one third of all the observed choices on four alternatives	\citep{GiarlottaPetraliaWatson2022a}.
Thus, the need for subclasses of limited attention that explain only a narrower fraction of choices becomes apparent.
Moreover, many models that retrieve choices by recurring to different behavioral explanations, such as \textit{rationality} \citep{Samuelson1938}, \textit{shortlisting} and \textit{list-rationality} \citep{Yildiz2016}, \textit{temptation} \citep{RavidStevenson2021}, and \textit{conspicuity} \citep{KibrisMasatliogluSuleymanov2021} are nested into the general limited attention framework, but they do not convey any information about the DM's attention filter.
Hence, we may ask which are their representations under limited attention, and the additional properties that the  attention filters must satisfy. 

In this note we represent each analyzed model as a special case of limited attention, and we identify the properties of the attention filters that distinguish the examined choice behavior.
By doing so, not only we propose an alternative narrative of these models, but we provide more restrictive patterns of limited attention.
Our investigation allows to retrieve information about the DM's consideration sets and preference from each observed choice; these welfare recommendations are compared with those provided by the equivalent frameworks.
Remarkably, there is always a DM's preference that is compatible with a model of choice and the corresponding subclass of limited attention. 
Finally, we examine the intersections among specifications, and we prove that each of them is independent of the others.  

\textit{Related literature.--} This paper contributes to the vast literature on deterministic choices.
According to the theory of revealed preferences pioneered by \cite{Samuelson1938}, preferences can be inferred from choice data. 
In this approach, rationality of choice functions is encoded by the notion of \textsl{rationalizability}, that is, the possibility to explain the DM's pick in each menu by maximizing a linear order over the alternatives.  
However, rationalizability does not justify many observed phenomena.   
Thus, starting from the seminal work of \cite{ManziniMariotti2007}, rationalizability has been weakened by several models of \textit{bounded rationality}, inspired by behavioral phenomena observed in experimental economics and psychology.
In our work we interpret rationality and many bounded rationality models as the observed behavior of a DM whose consideration satisfies specific features.
Moreover, the relationships among the specifications obtained from our analysis are explored.
In doing so, we build a bridge between limited attention and the other rationality/bounded rationality methods, and we offer a bird's eye on the representation, characterization, and nestedness of limited attention in choice theory. 

We also provide several results about the inference of unobserved consideration sets from the observed choice.
This problem has been already addressed in economic psychology and marketing, in which consideration sets are usually estimated by means of surveys \citep{PaulssenBagozzi2005,Suh2009}, and statistical inference applied on household panel data  \citep{VanNieropetal}.
In economic theory, starting from the paper of \cite{ManziniMariotti2014}, many solutions have been proposed to retrieve consideration probabilities from stochastic choice data.
In the deterministic case the elicitation of the DM's consideration  sets is still an open problem, because the same observed choice function can be explained by distinct attention filters.
We contribute to this research question by offering some results that show how the DM's preference and attention filter can be gathered from 
irrational features of choices.
Despite the adoption of consideration sets, the experimenter can always recover from data a DM's preference that fits a model of choice, and the equivalent specification of limited attention.  
 
The paper is organized as follows. 
Section~\ref{SECT:Preliminaries} contains preliminary definitions and results.
In Section~\ref{SECT:limited_attention} we analyze some  subclasses of limited attention that explain the same choice behavior justified by several models of choice that have been investigated in the literature.
Specifically, in Subsection~\ref{SECT:limited_attention}\ref{SUBSECT:optimal_limited_consideration} we present a special case of limited attention in which consideration costs affect the DM's choice, and we show that it is equivalent to a shortlisting procedure proposed by \cite{Yildiz2016}.
In Subsection~\ref{SECT:limited_attention}\ref{SUBSECT:Salient_limited_attention} the DM's attention is affected only by the most salient items she faces.
This framework justifies the same choice behavior explained by \cite{RavidStevenson2021} and \cite{KibrisMasatliogluSuleymanov2021}.
Moreover, we prove that rationalizability \citep{Samuelson1938} can be alternatively represented as a subclass of choices with limited attention in which DM's consideration sets are affected only by the item on top of her preference. 
In the method analyzed in Subsection~\ref{SECT:limited_attention}\ref{SUBSECT:Competitive_limited_attention} items pairwise compete for the DM's consideration.
This approach is equivalent to the model of \textit{list rational choices} prosed by \cite{Yildiz2016}.         
{For each of these novel specifications we display the properties of the associated attention filters, and we show how to partially identify them and the DM's preference from the observed violations of the contraction consistency axiom.
The independence of each subclass from the others, and their intersections are investigated in Section~\ref{SECT:relationships_subclasses}. 
Section~\ref{SECT:Concluding_remarks} collects some concluding remarks.
All the proofs are in the Appendix.

\section{Preliminaries}\label{SECT:Preliminaries}
In what follows, $X$ denotes the \textsl{ground set}, a finite nonempty set of alternatives.
 Any nonempty set $A \subseteq X$ is a \textsl{menu}, any nonempty set $B$ such that $B\subseteq A$ is a \textsl{submenu of}  $A$, and $\X = 2^X \setminus \{\es\}$ is the family of all menus.
A \textsl{choice function} on $X$ is a map $c \colon \mathscr{X}\rightarrow X$ such that $c(A)\in A$ for any $A\in\X$. 
A \textsl{choice correspondence} on $X$ is a map $\Gamma\colon \mathscr{X}\rightarrow  \mathscr{X}$ such that $\Gamma(A)\subseteq A$ for any $A\in\X$.
To simplify notation, we often omit set delimiters and commas: thus, $A \cup x$ stands for $A \cup \{x\}$, $A\setminus x$ stands for $A\setminus \{x\}$, $c(xy)=x$ for $c(\{x,y\})=x$, $\Gamma(xy)=x$ for $\Gamma(\{x,y\})=\{x\}$, etc.
In this work a choice function is interpreted as the observable outcome of an experiment in which the DM selects one alternative from each menu.
Instead, a choice correspondence, by mapping each menu to some submenu, resembles the unobserved collection of the DM's consideration sets.
A binary relation $\succ$ on $X$ is \textsl{asymmetric} if $x \succ y$ implies $\neg(y \succ x)$, \textsl{transitive} if $x \succ y \succ z$ implies $x \succ z$, and \textsl{complete} if $x \neq y$ implies $x \succ y$ or $y \succ x$ (here $x,y,z$ are arbitrary elements of $X$). 
A binary relation $\succ$ on $X$ is \textsl{acyclic} if $x_1 \succ x_2 \succ \ldots \succ x_n \succ x_1$ holds for no $x_1,x_2, \ldots, x_n \in X$, with $n \geqslant 3$.
A \textsl{strict partial order} is an asymmetric and transitive relation.
A \textsl{strict linear order} is an asymmetric, transitive, and complete relation.

Given an asymmetric relation $\succ$ on $X$ and a menu $A \in \X$, the set of \textsl{maximal} elements of $A$ is $\max(A,\succ)=\{x \in X : y \succ x \text{ for no } y \in A\}$.
 A choice function $c \colon \mathscr{X} \to X$ is \textsl{rationalizable} if there is a {strict} linear order $\rhd$ on $X$ such that, for any $A \in \mathscr{X}$, $c(A)$ is the unique element of the set $\max(A,\rhd)$; 
in this case we write $c(A) = \max(A,\rhd)$. 
\begin{definition}
A choice function $c \colon \mathscr{X} \to X$ is \textsl{rationalizable} \textit{(RAT)} if there is a {strict} linear order $\rhd$ on $X$ such that $c(A)=\max(A,\rhd)$ for any $A \in \mathscr{X}$.	
\end{definition}

Rationalizability of choice functions is characterized by the property of \textsl{Contraction Consistency} due to \cite{Chernoff1954},
 also called \textsl{Axiom}$\:\alpha$ by \cite{Sen1971}. 
This property states that if an item is chosen in a menu, then it is also chosen in any submenu:
\begin{description}
	\item[Axiom$\:\alpha$\,:\!]
  for all $A,B\in \X$ and $x \in X$, if $x \in A \subseteq B$ and $c(B)=x$, then $c(A)=x$. 
\end{description} 

Violations of Axiom$\:\alpha$ describe irrational features of the observed choice.  
To simplify our analysis, we introduce the following notation.

\begin{definition} \label{DEF:minimal_violations_of_alpha}
	For any choice function $c \colon \X \to X$, a \textsl{switch} is an ordered pair $(A,B)$ of menus such that $A \subseteq B$ and $c(A) \neq c(B) \in A$. 
A \textit{minimal switch} is a switch $(A,B)$ such that $\vert B\setminus A\vert=1$, or, equivalently, a pair $(A\setminus x,A)$ of menus such that $x\neq c(A) \neq c(A\setminus x)$.
\end{definition} 

Switches are violations of Axiom\,$\:\alpha$. 
A minimal switch $(A\setminus x,A)$ arises in a peculiar situation: if the DM chooses $c(A)$ from a menu $A$, and a $x$ distinct from $c(A)$ is removed from $A$, then the alternative $c(A\setminus x)$ selected from the smaller menu $A \setminus x$ is different from $c(A)$.
As argued in \cite{GiarlottaPetraliaWatson2022b}, for a finite ground set $X$ violations of Axiom\,$\alpha$ can always be reduced to minimal switches: 

\begin{lemma}[\citealp{GiarlottaPetraliaWatson2022b}] \label{LEMMA:minimal_violations_of_alpha}
Let $c \colon \X \to X$ be a choice function. 
For any switch $(A,B)$, there are a menu $C \in \X$ and an item $x \in X$ such that $A\subseteq C\setminus x \subseteq C \subseteq B$ and $(C\setminus x,C)$ is a minimal switch.
\end{lemma}

In view of Lemma~\ref{LEMMA:minimal_violations_of_alpha}, when we will analyze violations of Axiom\,$\alpha$ often we will refer only to minimal switches.


\section{Limited attention}\label{SECT:limited_attention}

In the influential approach of \cite{MasatliogluNakajimaOzbay2012} the observed dataset can be explained by the DM's limited attention. 
\begin{definition}[\citealp{MasatliogluNakajimaOzbay2012}]\label{DEF:CLA}
	A choice function $c\colon\X\to X$ is a \textsl{choice with limited attention} (\textsl{CLA}) if $c(A)=\max(\Gamma(A),\rhd)$ for all $A \in \X$, where:
	\begin{enumerate}[\rm(i)]
		\item 
		 $\rhd$ is a {strict} linear order  on $X$ (\textit{preference}), and 
		 \item $\Gamma \colon \X \to \X$ is a choice correspondence (\textit{attention filter}) such that for any $B \in \X$ and $x \in X$, $x \notin \Gamma(B)$ implies $\Gamma(B) = \Gamma(B \setminus x)$.
	 	\end{enumerate}
We say that $(\Gamma,\rhd)$ is an \textit{explanation by limited attention of} $c$.	 	
	 	\end{definition}

According to Definition~\ref{DEF:CLA}, when the DM faces a menu $A$, she first pays attention only to a submenu $\Gamma(A)$ of alternatives, and then she selects from it the best item $\max(\Gamma(A),\rhd)$  with respect to her preference $\rhd$.
The attention filter $\Gamma$ collects the DM's consideration sets, and it mirrors the characteristics of the DM's attention.
Indeed, if an item $x$ is not observed in $B$, i.e. $x\not\in\Gamma(B)$, then $\Gamma(B)$ must be equal to $\Gamma(B\setminus x)$, which is the DM's consideration set formed from the menu $B\setminus x$, in which $x$ has been removed from the available alternatives.

Given the weak constraints on the attention filter, Definition~\ref{DEF:CLA} is not restrictive, and it holds for a relatively large portion of choice functions defined on small ground sets.  
In fact, the authors point out that any choice function defined on a ground set of three items is a choice with limited attention. 
Moreover, \cite{GiarlottaPetraliaWatson2022a} prove that limited attention explains $324/864$ of all the choice functions on four alternatives.
On the other hand,  revealed preference analysis indicates that several models, which do not relate to the features of the DM's consideration, justify choice behaviors that are a portion of all the choices with limited attention defined on ground sets of arbitrary size. 
Thus, in the next subsections, by requiring that the attention filters satisfy additional properties, we introduce new subclasses of limited attention that reproduce behavioral characteristics of the DM's consideration.
We also show that each special case is an alternative representation of some rationality/bounded rationality method that appeals to selection biases different from those determined by limited attention.

\subsection{Optimal limited attention}\label{SUBSECT:optimal_limited_consideration}

In some situations, due to cognitive limitations and time constraints, the DM is not able to evaluate all the available alternatives, but she can select some items to consider, on the base of the cost of observing such alternatives.
This phenomenon has been originally analyzed by \cite{RobertsLattin1991}, who develop a framework of consideration set composition in which the DM compensates the expected benefits and costs of consideration of each brand.     
More recently, such process has been extensively discussed by \cite{CaplinDean2015} and \cite{CaplinDeanLeahy2019}, who assume that the DM pays attention to alternatives on the base of the trade-off between their expected value and the consideration cost.
The authors, by defining the DM's utility and information structure, create a setting that allows the experimenter to retrieve consideration costs and consideration sets from choice data.

These models, despite offering an accurate analysis of the DM's problem, rely on a high number of parameters, which may make their validation in lab experiments difficult. 
In light of this, we propose a simple specification of limited attention in which consideration costs matter.
Indeed, we assume that the DM is able to rank some alternatives according to their observational burden, and from each menu she pays attention to the items with the lowest consideration cost.
 Thus, the associated attention filter replicates her concerns about \textit{optimal} attention.

\begin{definition}\label{DEF:rational_limited_consideration}
	A choice function $c\colon \X\to X$ is a \textit{choice with optimal limited attention (COLA)}  if $c(A)=\max(\Gamma^{\,opt}(A),\rhd)$ for any $A\in\X$, where
	\begin{enumerate}[\rm(i)]
		\item $\rhd$ is a {strict} linear order on $X$, and
	\item $\Gamma^{\,opt}\colon \X \to \X$ is a choice correspondence (\textsl{optimal attention filter}) such that, for any $B\in\X$ and $x,y\in X$, we have:
		\begin{enumerate}
			 \item if $x \in B$ and $x \notin  \Gamma^{\,opt}(B)$, then $\Gamma^{\,opt}(B)\supseteq \Gamma^{\,opt}(B\setminus x);$
				\item if $x\in B$ and $y \in{(\Gamma^{\,opt}(xy) \cap \Gamma^{\,opt}(B\setminus x))}$, then $y \in \Gamma^{\,opt}(B)$; 
				\item $\Gamma^{\,opt}(B)\setminus x \sbs \Gamma^{\,opt}(B\setminus x)$.
	\end{enumerate}
	\end{enumerate}
We say that $(\Gamma^{\,opt},\rhd)$ is an \textit{explanation by optimal limited attention of} $c$.
\end{definition}

Condition (ii)(a) of Definition~\ref{DEF:rational_limited_consideration}  imposes that the removal of an unobserved item from some menu cannot increase the amount of alternatives considered by the DM.
Since $x$ is not observed, its consideration cost must be higher than some alternative $y$ considered in $B$.
If another item $z$ is not observed in $B$ because the cost of considering it is higher than $x$, then the DM does not pay attention to $z$ in $B\setminus x,$ because its consideration cost is higher than $y$.
 
Condition (ii)(b) requires that if the DM pays attention to $y$ when she faces the menus $xy$ and $B\setminus x$, then she must consider $y$ from the menu $B$, obtained by adding $x$.
Since $y$ is considered in $xy$ and $B\setminus x$, we can conclude that its consideration cost is equal or lower than $x$ and any alternative in $B\setminus x$.
Therefore, we expect that the DM pays attention to $y$ from $B$.
Condition (ii)(c) tells that if an item $x$ is added to the menu $B\setminus x$, then alternatives distinct from $x$ and considered in $B$ must have been considered in $B\setminus x$.
If the DM pays attention to some alternative $z$ from the menu $B$, then its consideration cost is equal or lower than $x$ and any item in $B\setminus x$.  
Note that conditions (ii)(a) and {(ii)(c)} imply that any optimal attention filter is an attention filter, and, as a consequence, any choice with  optimal limited attention is a choice with limited attention.
{\begin{remark}
Any  restriction on $\Gamma^{\,opt}$ is not implied by the others.
To prove this fact, in the Appendix we exhibit three choice correspondences, each one satisfying two of the conditions (ii)(a), (ii)(b), and (ii)(c) of Definition~\ref{DEF:rational_limited_consideration}, but not the third one. 
\end{remark}}
Optimal limited attention explains the same choice behavior justified by the model of \cite{Yildiz2016}, in which the DM selects the alternative that survives the sequential applications of two distinct criteria. 

\begin{definition}[\citealp{Yildiz2016}]\label{DEF:shortlisting}
	A choice function $c\colon\X\to X$ is a \textit{shortlisting (SL)} if there are a {strict} partial order $>$ and a {strict} linear order $\rhd$ on $X$ such that, for any $A \in \X$, we have $c(A)=\max(\max(A, >), \rhd)$.
	We say that the pair $(>,\rhd)$ \textit{describes} $c$.
\end{definition}

This approach is a special case of the procedures discussed by \cite{ManziniMariotti2007}, \cite{Tyson2013}, and \cite{DuttaHoran2015}.
Indeed, we have:

\begin{theorem}\label{THM:rational_attention_equivalent_shortlisting}
	A choice function is a shortlisting if and only if it is  a choice with optimal limited attention.
	\end{theorem}

Thus, a shortlisting can be also interpreted as the observed choice of a DM whose attention is limited, and optimal, since it respects a ranking among the alternatives describing their consideration cost.
The next result sheds light on the information about the DM's optimal attention filter and preference that can be retrieved from the violations of Axiom\,$\alpha$ detected in the dataset.

\begin{lemma}\label{LEMMA:necessary_conditions_of_RLC}
	 Let $c\colon \X\to X$ be a choice with optimal limited attention, and assume that there are $A\in\X$ and $x\in A$ such that $x\neq c(A)\neq c(A\setminus x)$.
	Then $x> c(A\setminus x)$, $x\in\max(A,>)$, and $c(A\setminus x)\rhd c(A)\rhd x$ hold for any pair $(>,\rhd)$ that describes $c$.
	Moreover, $x\in\Gamma^{\,opt}(A), c(A\setminus x)\not\in\Gamma^{\,opt}(A),  c(A\setminus x)\not\in\Gamma^{\,opt}(c(A\setminus x)\,x), c(A)\in\Gamma^{\,\,opt}(c(A)\,x),$ and $c(A\setminus x)\rhd c(A)\rhd x$ hold for  any explanation by optimal limited attention $(\Gamma^{\,\,opt},\rhd)$ of $c$.
\end{lemma}

Lemma~\ref{LEMMA:necessary_conditions_of_RLC} shows that shortlisting and optimal limited attention reveal similar information about the DM's preference, and the items discarded in the first stage of her decision.
According to the procedure of \cite{Yildiz2016}, a minimal switch $(A\setminus x,A)$ implies that $x$ dominates $c(A\setminus x)$ in the DM's criterion adopted first, and it is among the items shortlisted in $A$.
Moreover, $c(A\setminus x)$ is preferred to $c(A)$, which in turn is preferred to $x$.
However, if the experimenter adopts a framework of optimal limited attention, he can partially identify the DM's attention filter from observed violations of contraction consistency axiom. 
Indeed, he can conclude that the DM pays attention to $x$ from $A$, and the removal of $x$ from that menu induces her to consider and select the item $c(A\setminus x)$, which has not been considered in $A$.
Moreover, in the pairwise comparison  with $x$, the DM does not pay attention to $c(A\setminus x)$.
The shape of the these three consideration sets are determined by the consideration cost of $x$, which is lower than $c(A\setminus x)$.
Finally, note that $c(A)$ is considered in $c(A)x$.
Since the DM pays attention to $c(A)$ and $x$ from $A$, either $c(A)$ and $x$ have the same consideration cost, or the DM is not able to compare their consideration burden.

{The welfare recommendations of Lemma~\ref{LEMMA:necessary_conditions_of_RLC} suggest that a given preference can be utilized to describe a choice function, or in a explanation by optimal limited attention of it .}
Moreover, we may expect that the consideration sets shaped by an optimal attention filter are also the items maximal with respect to some {strict} partial order.    
These facts are confirmed by the following lemma, and the corollary that goes with it. 

\begin{lemma}\label{LEM:partial_order_equals_optimal_attention_filter}
	The following statements are equivalent for a choice correspondence $\Gamma\colon\X\to \X$ on $X$:
	\begin{itemize}
		\item[\rm(i)] $\Gamma(A)=\max(A,>)$ for some {strict} partial order $>$ on $X$;
		\item[\rm(ii)] $\Gamma$ is an optimal attention filter.
	\end{itemize}
\end{lemma}

\begin{corollary}\label{COR:Optimal_limited_attention_shortlisting_same_preference}
	Let $\rhd$ be a {strict} linear order on $X$.
The following statements are equivalent for a choice function $c\colon \X\to X$:
\begin{itemize}
	\item[\rm(i)]  there is a {strict} partial order $>$ on $X$ such that $(>,\rhd)$ describes $c$; 
	\item[\rm(ii)] there is an optimal attention filter $\Gamma^{\,opt}\colon \X \to \X$ such that $(\Gamma^{\,opt},\rhd)$ is an explanation by optimal limited attention of $c$.
\end{itemize}
\end{corollary}

Note also that, as for attention filters, the properties of an optimal attention filter do not involve the DM's preference.
This detachment is a desirable feature of a limited attention specification, since it assumes that the DM's consideration is not affected by her taste.
However, such separation is apparent.
Indeed, the proof adopted by \cite{MasatliogluNakajimaOzbay2012} in the characterization of a choice with limited attention, and the proof of Theorem~\ref{THM:rational_attention_equivalent_shortlisting} indicate that an attention/optimal attention filter can be always defined by using some strict linear order that extends an asymmetric and acyclic revealed preference. 


\subsection{Salient limited attention}\label{SUBSECT:Salient_limited_attention}
In this subsection we discuss a special case of limited attention, in which the DM's consideration is influenced by \textsl{salience} of items, intended as the relatively extreme position of alternatives in her judgment.
The effects of salience on individual choices has been analyzed by \cite{BordaloGennaioliShleifer2012,BordaloGennaioliShleifer2013}, who argue that the DM's evaluation of goods whose attributes are far from the average can be distorted.
Recently, also \cite{Lanzani2022} proposes an axiomatization of salience theory under uncertainty.
These approaches analyze the distortions of the DM's preference determined by extreme features of available options, but they do not elaborate on the consequences of salience on attention. 

However, new findings in psychology \citep{ParrandFriston2017,ParrandFriston2019} report that there is an intimate relationship between these two facets of individual perception.  
Thus, \cite{GiarlottaPetraliaWatson2022b} propose a subclass of limited attention in which only salient items affect the DM's consideration.

\begin{definition}\label{DEF:CSLA} \rm 
	A choice function $c \colon \X \to X$ is a \textsl{choice with salient limited attention} \textsl{(CSLA)} if $c(A) = \max(\Gamma^{\,s}(A),\rhd)$ for all $A \in \X$, where
\begin{enumerate}[\rm(i)]
	\item $\rhd$ is a {strict} linear order on $X$, and  
	\item $\Gamma^{\,s} \colon \X \to \X$ is a choice correspondence (\textsl{salient attention filter}) such that, for any $B\in\X$ and $x\in B$, if $x \neq \min(B,\rhd),\max(\Gamma^{\,s}(B),\rhd)$, then $\Gamma^{\,s}(B) \setminus x = \Gamma^{\,s}(B \setminus x)$.  
\end{enumerate}
We say that $(\Gamma^{\,s},\rhd)$ is an \textit{explanation by salient limited attention of $c$}.
\end{definition}

Differently from the general framework of \cite{MasatliogluNakajimaOzbay2012}, in which  a variation in the DM's consideration set occurs only if some observed item is removed from the menu, condition (ii) of  Definition~\ref{DEF:CSLA} states that for a choice with salient limited attention the same anomaly is caused only by the removal either of the best considered alternative, or of the worst option in the menu.
A salient attention filter that, paired with some { 
strict} linear order, justifies a choice, is not necessarily an attention filter.
However, for any choice with salient limited attention there exists a suitable salient attention filter that is also an attention filter.
\begin{lemma}\label{LEM:existence_salience_attention_filter_attention_filter}
For any choice  with salient limited attention $c\colon\X\to X$ there is an explanation by salient limited attention $(\Gamma^{\,s},\rhd)$ such that $\Gamma^{\,s}$ is an attention filter.
	
\end{lemma}

The pattern discussed in Definition~\ref{DEF:CSLA} has the same predictive power of two models, which reproduce the effects of temptation and reference dependence on individual choices.

	\begin{definition}[\citealp{RavidStevenson2021}]\label{DEF:temptation}
		A choice function $c\colon \X\to X $ has a \textit{general-temptation representation (GTR)} $(u,v,W)$ if functions $u\colon X\to \mathbb{R}$, $v \colon X\to \mathbb{R}$, and $W\colon\mathbb{R}^{2}\to~\mathbb{R}_{+}$ exist such that $W$ is strictly increasing in both coordinates and $$c(A)=\arg \max_{x\in A}W\left(u(x), v(x)-\max_{y\in A}v(y)\right)$$ holds for any $A\in \X$.
\end{definition}

According to this approach, in any menu the DM selects the item that brings her the best trade-off between her utility, described by $u$, and the self-control cost $v$ of resisting the most tempting feasible option.
Note that in a general temptation representation the DM's consideration does not play any role.
The same is true for the following choice procedure.
\begin{definition}[\citealt{KibrisMasatliogluSuleymanov2021}]\label{DEF:endogenous_reference_representation}
	A choice function $c\colon\X\to X$   \textit{admits a conspicuity based endogenous reference representation} (\textit{CER})  if there exist a family of injective real valued functions $U=\{U_{z}\}_{z\in X}$ and a {strict} linear order $\gg$ on $X$ such that 
	
	$$c(A)=\argmax_{x\in A}U_{\max(A,\gg)}(x).$$

We call $(\gg,\{U_{z}\}_{z\in X})$ a \textit{conspicuity based endogenous reference representation of} $c$.	 
\end{definition}

Definition~\ref{DEF:endogenous_reference_representation} states that in any menu $A$ the DM's perception is captured by the most conspicuous item $\max(A,\gg)$, which moves her to adopt the utility $U_{\max(A,\gg)}$ in her selection.
The choice behavior explained by general temptation and conspicuity based endogenous reference representation is also justified by salient limited attention.
Indeed, we have:

{\begin{theorem}
\label{THM:CSLA_equivalent_representation}
The following are equivalent for a choice function $c$:
\begin{enumerate}[\rm(i)] 
	\item $c$ has a general temptation representation;
	\item $c$ admits a conspicuity based endogenous reference representation;
	\item $c$ is a choice with salient limited attention.
	\end{enumerate}
\end{theorem}     

Moreover, salient limited attention allows the experimenter also to retrieve information about the DM's consideration sets. 

\begin{lemma}~\label{LEM:CSLA_partial_identification}
  Let $c\colon \X\to X$ be a choice with salient limited attention,
and assume that there exist $A\in\X$ and $x\in A$ such that $x\neq c(A)\neq c(A\setminus x)$.
Then, we have that $v(x)=\max_{y\in A} v(y)$, and  either $u(c(A))> u(c(A\setminus x))$, $v(c(A))< v(c(A\setminus x))$ or $u(c(A))< u(c(A\setminus x))$, $v(c(A)) > v(c(A\setminus x))$ hold  for any general-temptation representation $(u,v,W)$ of $c$.
 Moreover, $x=\max(A,\gg)$, and $c(A)=\argmax_{y\in A} U_x(y) $ hold for any conspicuity based endogenous reference representation $(\gg,\{U_{z}\}_{z\in X})$ of $c$.
Finally, $\Gamma^{\,s}(A\setminus x)\neq \Gamma^{\,s}(A)\setminus x$, $x\in \Gamma^{\,s}(A)$, and $x=\min(A,\rhd)$ hold for any explanation by salient limited attention $(\Gamma^{\,s},\rhd)$ of $c$.

\end{lemma}

 Lemma~\ref{LEM:CSLA_partial_identification} also compares the welfare indications obtained from data by salient limited consideration with those conveyed by the other two models of choice.
Indeed, if the experimenter assumes that the DM's choice is influenced by temptation, he can deduce from a minimal switch  $(A\setminus x,A)$ 
that $x$ is the most tempting alternative in $A$, and that 
$c(A)$ is more tempting than $c(A\setminus x)$, but it brings lower utility than $c(A\setminus x)$, or vice versa.
According to the conspicuity based endogenous reference representation, $x$ is more conspicuous than any other item in $A$, and it induces the DM to adopt the utility $U_{x}$ to finalize her selection from $A$.     
Under the hypothesis of salient limited attention, the experimenter infers that the subset of alternatives observed from $A$ is distinct from the consideration set determined by the DM when she faces the menu $(A\setminus x)$.
Moreover he can deduce not only that the DM pays attention to $x$ from $A$, but also that $x$ is the least preferred item in that menu.

Remarkably, it is always possible to identify a DM's preference that is compatible with any method among those presented in Definitions \ref{DEF:temptation}, \ref{DEF:endogenous_reference_representation}, and \ref{DEF:CSSLA}.
Before formally introducing this fact, we need some additional notation.
Given a {strict} linear order $\gg$ on $X$, we denote by $x^{\gg}_{\vert X\vert}$ the item $\min(X,\gg),$ and by $x^{\gg}_{\vert X\vert-1}$ the item $\min(X\setminus \min(X,\gg),\gg).$
Moreover, we denote by $\widetilde{P}$ the binary relation on $X$ defined by $x\widetilde{P}y$ if there is a menu $A$ such that $x\in A$ and $(A\setminus y, A)$ is a switch.
Finally we denote by $\rhd^{\widetilde{P},\gg}$ a {strict} linear order (if any) such that $\widetilde{P}\subseteq\rhd^{\widetilde{P},\gg},$ and  $x^{\gg}_{\vert X\vert-1}\rhd^{\widetilde{P},\gg}\; x^{\gg}_{\vert X\vert}$ if  $c\left(x^{\gg}_{\vert X\vert-1}x^{\gg}_{\vert X\vert}\right)=x^{\gg}_{\vert X\vert-1},$ or $x^{\gg}_{\vert X\vert}\rhd^{\widetilde{P},\gg}\;x^{\gg}_{\vert X\vert-1}$ if  $c\left(x^{\gg}_{\vert X\vert-1}x^{\gg}_{\vert X\vert}\right)=x^{\gg}_{\vert X\vert}.$
The following result holds.

\begin{lemma}\label{LEM:common_preference_among_models}
Assume that $c\colon \X\to X$ is a choice with salient limited attention.
Then there is a {strict} linear order $\rhd$ on $X$ such that
\begin{itemize}
 \item[\rm(i)] for some  general-temptation representation $(u,v,W)$ of $c$, and any $x,y\in X$, $u(x)>u(y)$ holds if and only if $x\rhd y$ is true, 
\item[\rm(ii)] for some conspicuity based endogenous reference representation  $(\gg,\{U_{z}\}_{z\in X})$ of $c$, some $z\in X\setminus x^{\gg}_{\vert X\vert}$, and  any  $x,y\in X$, $U_{z}(x)>U_{z}(y)$ holds if and only if $x\rhd y$ is true, and
 \item[\rm(iii)] 
 $(\Gamma^{\,s},\rhd)$ is an explanation by salient limited attention of $c$ for some salient attention filter $\Gamma^{s}\colon \X\to\X$,
\end{itemize}
Moreover, $\rhd\equiv\rhd^{\widetilde{P},\gg}.$
\end{lemma}

Lemma~\ref{LEM:common_preference_among_models} can be used by the experimenter to detect a preference that belongs to some explanation by salient limited attention, and that is represented by the utility functions respectively of some general-temptation representation and conspicuity based endogenous reference representation of the observed choice.
Such preference, and the utility that represents it, always drive the DM's selection from some menu.
This fact, which is evident for temptation and salient limited attention, in the setting of \cite{KibrisMasatliogluSuleymanov2021} is determined by the requirement that the utility that represents $\rhd^{\widetilde{P},\gg}$ cannot be indexed by the less conspicuous item in $X$ according to $\gg$, namely $x^{\gg}_{\vert X\vert}.$ 
Salient limited attention offers also a novel representation of rationalizable choice functions.
Indeed, we single out a special case of the framework discussed in Definition~\ref{DEF:CSLA}, in which the  DM's consideration is affected only by the best considered item.
\begin{definition} \label{DEF:CSSLA} \rm 
	A choice function $c \colon \X \to X$ is a \textsl{choice with selective salient limited attention} \textsl{(CSSLA)} if $c(A) = \max\left(\Gamma^{\,sl}(A),\rhd\right)$ for any $A \in \X$, where
\begin{enumerate}[\rm(i)]
	\item $\rhd$ is a {strict} linear order on $X$, and   
	\item  $\Gamma^{\,sl} \colon \X \to \X$ is a choice correspondence (\textsl{selective salient attention filter}) such that, for all $B \in \X$ and $x \in X$, if $x \neq\max\left(\Gamma^{\,sl}(B),\rhd\right)$, then $\Gamma^{\,sl}(B) \setminus x = \Gamma^{\,sl}(B \setminus x)$.
\end{enumerate}
We say that $(\Gamma^{sl},\rhd)$ is an \textit{explanation by selective salient limited attention of $c$}.
\end{definition}

According to Definition~\ref{DEF:CSSLA}, a variation in the DM's consideration set occurs only if the best item, among those observed, is removed from the menu.
Selective salient attention captures rationalizable choice behavior.

\begin{theorem}\label{THM:CSSLA_alpha}
A choice function is rationalizable if and only if it is a choice  with selective salient limited attention.
\end{theorem}

The result above provides a representation under limited attention of rationalizable choices in which some alternatives may not be observed.  
Indeed, the proof of Theorem~\ref{THM:CSSLA_alpha} shows that for any choice function satisfying Axiom\;$\alpha$ there is an explanation by selective salient limited attention whose attention filter excludes some alternatives from some menus.


\subsection{Competitive limited attention}\label{SUBSECT:Competitive_limited_attention}

A massive strand of the marketing literature reports a fierce competition among products and brands to gain the DM's attention.  
Indeed, as a response to individual tastes, aggressive advertising, and time pressure, products that receive attention tend to  substitute competitors in the consumer's consideration.
In this respect, \cite{FaderMcAlister1990} develop and compute parametric estimations of a model in which consumers do not consider brands that are not on promotion.
\cite{AllenbyGinter1995} find that in-store merchandising increases the probability that products with the same brand names are jointly considered, and  reduces the likelihood that  alternative brands are taken into account by households.
In the work of \cite{TeruiBanAllenby2021} there is evidence that advertising raises the probability that a given item is included in the consumer's consideration set, and that other alternatives are ruled out.
We reproduce the effects of marketing strategies on individual choice, by assuming that some brands and products may exclude other options from the DM's attention.

\begin{definition} \label{DEF:list_attention}
	A choice function $c\colon\X\to X$ is a \textit{choice with competitive limited attention (CCLA)} if $c(A)=\max(\Gamma^{\,co}(A),\rhd)$ for any $A\in\X$, where:
	\begin{enumerate}[\rm(i)]
		\item $\rhd$ is a {strict} linear order on $X$ (\textsl{preference}), and
		\item $\Gamma^{\,co}: \X \to \X$ is a choice correspondence (\textit{competitive attention filter}) such that the following properties hold for any $A\in\X$:
		\begin{enumerate}
			\item if $x \in A$ and $x \notin  \Gamma^{\,co}(A)$, then $\Gamma^{\,co}(A)=\Gamma^{\,co}(A\setminus x)$;
		\item for all $x,y \in X$ such that  $y =\max(A,\rhd)$, and $x=\max(\Gamma^{\,co}(A\setminus y),\rhd)$, then 
		$$
		 y \in \Gamma^{\,co}(xy) \iff y \in \Gamma^{\,co}(A).
		$$
	\end{enumerate}
	\end{enumerate}
We say that $(\Gamma^{\,co},\rhd)$ is an \textit{explanation by competitive limited attention of} $c$.
\end{definition}

Condition (ii)(a) imposes that an ignored item does not influence the DM's consideration set when it is removed from the menu, and it shows that competitive limited attention is a subclass of the general paradigm of \cite{MasatliogluNakajimaOzbay2012}. 
However, competitive attention filters satisfy an additional property, described in (ii)(b), according to which if an item $y$ is on top of the DM's preference in a menu $A$, then it is considered in $A$ if and only if it is considered when $c(A\setminus y)$ is the only other available option.
These two options \textit{compete} for the DM's attention, and if 
$c(A\setminus y)$  overcomes $y$ in a pairwise comparison, then $c(A\setminus y)$ will always shade $y$ in the DM's mind, unless an option better than $y$ is added to the menu. 
The subclass of limited attention described in Definition~\ref{DEF:CLA} explains the same choice data justified by \textsl{list-rational} choices, introduced by \cite{Yildiz2016}.

\begin{definition}[\citealp{Yildiz2016}] \rm \label{DEF:list}
    A choice function $c\colon\X\to X$ is \textit{list rational} (\textsl{LR}) if there is a {strict} linear order $\mathbf{f}$ (a \textit{list}) on $X$ such that for each menu $A \in\X$, we have $c(A)=c\big(\{c(A\setminus \max(A,\mathbf{f})),\max(A,\mathbf{f})\}\big)$.
    We say the $\mathbf{f}$ \textit{list rationalizes} $c$.\footnote{According to \cite{Yildiz2016} the item $\max(A,\mathbf{f})$ is the \textit{last alternative in A according to} $\mathbf{f}$.
    Thus, for the author a list $\mathbf{f}$ is to read from bottom to top.
    {For instance, given the set $X=\{x,y,z\}$,  $x\,\mathbf{f}\,y\,\mathbf{f}z$ means that $z$ is the first item of the list $\mathbf{f}$, $y$ is the second, and $x$ is the last one.}
   }
	\end{definition}

According to this procedure the DM picks in each menu the option that survives from a binary comparison of alternatives on a list.
A list rational choice function can be equivalently interpreted as the selections of a DM influenced by marketing strategies that aim to exclude some products from her attention.

\begin{theorem} \label{THM:list_rationalizable}
	A choice function is list rational if and only if it is a choice with competitive limited attention.
\end{theorem}

The experimenter can retrieve information about the DM's consideration sets, and her preference from violations of Axiom\;$\alpha$ displayed by a choice with competitive limited attention.
Indeed, we have:

\begin{lemma}\label{LEMMA:necessary_conditions_of_CCLA}
Let $c\colon \X\to X$ be a choice with competitive limited attention.
If there are $A \in \X$ and $x\in A$ such that $x\neq c(A)\neq c(A\setminus x)$, then $c(A)\,\mathbf{f}x$ for any $\mathbf{f}$ that list rationalizes $c$. 
Moreover, $\Gamma^{\,co}(A\setminus x)\neq\Gamma^{\,co}(A), $ $x\in \Gamma^{\,co}(A)$, and $c(A)\rhd x$ hold for any explanation by competitive limited attention $(\Gamma^{\,co},\rhd)$ of $c$.
If there are $A\in\X$, and $x,y\in A$ such that either $x=c(xy)=c(A\setminus y)\neq c(A)=y$, or $y=c(xy)\neq c(A\setminus y)=c(A)=x$, then $x\,\mathbf{f}y$ for any $\mathbf{f}$ that list rationalizes $c$.  
Furthermore, for  any explanation by competitive limited attention $(\Gamma^{\,co},\rhd)$ of $c$, we have that $z\rhd y$ for some $z\in A	\setminus y$.
 \end{lemma}

According to list rationality, a switch $(A\setminus x,A)$ implies that $x$ must be  followed by $c(A)$ in any list that list rationalizes the DM's choice. 
Moreover, when the DM selects $x$ from $xy$ and $A\setminus y$, and  $y$ is picked from $A$, it can be inferred that $x$ always follows $y$ in the DM's mind.
The experimenter knows that $x$ comes after $y$ also when $y$ is selected from $xy$, and $x$ is picked from $A\setminus y$ and $A$.
 For a choice with competitive limited attention, if removing $x$ from a menu $A$ in which $y$ is selected causes a minimal switch, then we can infer that the DM pays attention to $x$ at $A$, the set of items considered from $A\setminus y$ is different from the DM's consideration set at $A$, and she prefers $y$ to $x$.
 Moreover, if an item $x$ is picked from binary comparison with $y$, and from a menu $A\setminus y$ not containing $y$, but $y$ is selected from $A$, we conclude that there must be an item $z$ in $A\setminus y$ that is preferred to $y$.
The same can be inferred when $y$ is selected from $xy$, but $x$ is chosen from $A\setminus y$, and $A$. 

Lemma~\ref{LEMMA:necessary_conditions_of_CCLA} hints that the list adopted in \cite{Yildiz2016} is behaviorally equivalent to the DM's preference of a choice with competitive limited attention.
Indeed, the proof of Theorem~\ref{THM:list_rationalizable} implies that a list rationalizes the observed choice if and only if it belongs to some explanation by competitive limited attention of data.

\begin{corollary}
	\label{COR:list_competitive_limited_attention_same_preference}
	Let $\rhd$ be a {strict} linear order on $X$.
The following statements are equivalent for a choice function $c\colon \X\to X$:
\begin{itemize}
	\item[\rm(i)]  $\rhd$ list rationalizes $c$;
	\item[\rm(ii)] there is a competitive attention filter $\Gamma^{\,co}\colon \X \to \X$ such that $(\Gamma^{\,co},\rhd)$ is an explanation by competitive limited attention of $c$.
	\end{itemize}
\end{corollary}

Theorem~\ref{THM:list_rationalizable}, Lemma~\ref{LEMMA:necessary_conditions_of_CCLA}, and Corollary~\ref{COR:list_competitive_limited_attention_same_preference}  conclude our comparison between specifications of limited attention and models of choice. 
The main findings of our analysis are summarized in Table~\ref{TABLE:summary_limited_consideration}.  

\begin{table}[H]
\begin{center}
\begin{tabular}{|c|c|c|}
\hline
Model & Limited attention specification & Attention filter \\
\hline
 SL & COLA & $\Gamma^{\,opt}$\\
\hline
GTR, CER & CSLA & $\Gamma^{\,s}$\\
\hline
RAT & CSSLA & $\Gamma^{\,sl}$\\
\hline
LR & CCLA & $\Gamma^{\,co}$ \\
\hline
\end{tabular}
\end{center}
\caption{Each model of choice in the first column is equivalent, in terms of observed behavior, to one of the special cases of limited attention listed in the second column, and discussed in Section~\ref{SECT:limited_attention}.
In the third column we display, for each row, the attention filter associated to the corresponding specification.}
\label{TABLE:summary_limited_consideration}
\end{table}

In the next section we investigate some relations between the subclasses of limited attention displayed in Table~\ref{TABLE:summary_limited_consideration}.
Moreover, we will analyze the connection with other patterns of limited consideration discussed in the literature.


\section{Relationships between subclasses}\label{SECT:relationships_subclasses}

The existence of many specifications of limited attention necessitates an understanding of their relationships.
Thus, in Subsection~\ref{SECT:relationships_subclasses}\ref{SUBSECT:independence_of_spefication}, by analyzing the intersections between subclasses, we show that each one is independent of the others, and we exhibit some attention filters that satisfy several properties among those listed in Definitions~\ref{DEF:rational_limited_consideration}, \ref{DEF:CSSLA}, and \ref{DEF:list_attention}.
Moreover, in Subsection~\ref{SECT:relationships_subclasses}\ref{SUBSECT:path_independent_consideration} we compare competitive limited attention with the methods proposed in \cite{LlerasMasatliogluNakajimaOzbay2017} and \cite{LlerasMasatliogluNakajimaOzbay2021}. 
Before doing so, we present some revealed preference relations, whose properties characterize the subclasses of limited attention analyzed in the previous sections.

\begin{definition}[\citealp{Yildiz2016}]\label{DEF:revealed_shortlisting}
   For any $x,y\in X$, we say that $xR^cy$ if and only if at least one of the three following conditions hold:
    \begin{itemize}
    	\item[\rm(i)] there is $A \in \X$, such that $x=c(A \cup y)$ and $x\neq c(A)$;
    	\item[\rm(ii)] there is $A \in\X$, such that $y=c(A \cup x)$ and $x=c(xy)$;
    	\item[\rm(iii)] there is $A \in\X$, such that $y \neq c(A \cup x)$, $y=c(xy)$, and $y=c(A)$.
    \end{itemize}
\end{definition}

\begin{definition}[\citealt{GiarlottaPetraliaWatson2022b}]\label{DEF:revealed_salience}
For any $x,y\in X$, we say that $x\vDash_c y$ if there is there is a menu $A\in\X$ such that $x,y\in A$ and $(A\setminus x, A)$ is a minimal switch. 
\end{definition}

\begin{definition}[\citealp{Yildiz2016}]\label{DEF:revealed_list}\label{revealed_to_follow}
   For any $x,y\in X$, we say that $xF_c y$ (\textsl{x is revealed to follow to y}) if and only if at least one of the three following conditions hold:
    \begin{itemize}
    	\item[\rm(i)] there is $A \in \X$ such that $x=c(A \cup y),$ and $y=c(xy)$;
    	\item[\rm(ii)] there is $A\in\X$ such that $x=c(A \cup y),$ and $x \neq c(A)$;
    	\item[\rm(iii)] there is $A \in\X$ such that $x\neq c(A \cup y),$ $x=c(xy),$ and $x=c(A)$.
    \end{itemize}
\end{definition}

According to \citet[Proposition $2$ and Corollary $1$]{Yildiz2016}, acyclicity and asymmetry of $R^c$ and $F_c$ respectively characterize shortlistings and list rational choice functions. 
Moreover, \citet[Theorem $3$]{GiarlottaPetraliaWatson2022b} show that the asymmetry of $\vDash_c$ is a necessary and sufficient condition of any choice that admits a conspicuity based endogenous reference representation. 
Thus, by Theorems~\ref{THM:rational_attention_equivalent_shortlisting}, \ref{THM:CSLA_equivalent_representation}, and \ref{THM:list_rationalizable} we have:

\begin{corollary}\label{COR:optimal_limited_attention_characterization_revealed_preference}
A choice function is a choice with optimal limited attention if and only if $R^c$ is asymmetric and acyclic;	
\end{corollary} 

\begin{corollary}\label{COR:salient_limited_attention_characterization_revealed_preference}
A choice function is a choice with salient limited attention if and only if $\,\vDash_c$ is asymmetric;	
\end{corollary}

\begin{corollary}\label{COR:competitive_limited_attention_characterization_revealed_preference}
	A choice function is a choice with competitive limited attention if and only if  $F_c$ is asymmetric and acyclic.
\end{corollary}

In the next subsections we use these results to test whether a choice dataset belongs to some specification of limited attention.

\subsection{Independence and intersections}\label{SUBSECT:independence_of_spefication}
First, we exhibit some choice functions that belong to a subclass of limited attention, but not to the others.

	\begin{description}
		\item [\textbf{COLA, not CSLA, not CCLA}]Let $c_1\colon \X\to X$ be the choice function on $X=\{v,w,x,y,z\}$ defined by 
		 $$ vwxy\underline{z} \;\; vwx\underline{y} \;\; v\underline{w}xz \;\;  \underline{v}wyz \;\;  vxy\underline{z} \;\; wxy\underline{z} \;\;    v\underline{w}x \;\; \underline{v}wy \;\;   v\underline{w}z \;\; 
        vx\underline{z} \;\; vx\underline{y} \;\; \underline{v}yz \;\;
         $$
        $$
       wx\underline{y} \;\;\underline{w}xz \;\; wy\underline{z} \;\;  xy\underline{z} \;\; v\underline{w} \;\; v\underline{x} \;\;  \underline{v}y \;\; \underline{v}z \;\;    \underline{w}x \;\; w\underline{y}  \;\; \underline{w}z \;\;  x\underline{y} \;\; x\underline{z} \;\; y\underline{z}\footnote{In each menu, the underlined item is the one selected according to $c$.
       Moreover, we omit menus containing only one alternative, since they do not provide any relevant information.}.  
        $$

		The reader can check that the pair $(P,\rhd)$ such that $xPv, yPw$, and $w \rhd v \rhd z \rhd y\rhd x$ describes $c_1$.
		Thus, by Theorem~\ref{THM:list_rationalizable} $c_1$ is a choice with optimal limited attention, and by Lemma~\ref{LEM:partial_order_equals_optimal_attention_filter} and Corollary~\ref{COR:Optimal_limited_attention_shortlisting_same_preference} the pair $(\Gamma^{\,opt},\rhd)$ such that $\Gamma^{\,opt}(A)=\max(A,P)$ is an explanation by optimal limited attention of $c_1$.
		The selections $vwxy\underline{z}$, $\underline{v}wyz,$ and $v\underline{w}xz$ yield, by Definition~\ref{DEF:revealed_salience}, $x \vDash_{c_1} y$ and $y \vDash_{c_1} x$.
		Thus, by Corollary \ref{COR:salient_limited_attention_characterization_revealed_preference} $c_1$ is not a choice with salient limited attention.
		Moreover, the selections $\underline{v}wyz$, $v\underline{w}z$, $vwx\underline{y}$, and $\underline{v}y$ yield, by Definition \ref{DEF:revealed_list}, $v F_{c_1} y$, and $y F_{c_1} v$. Corollary~\ref{COR:competitive_limited_attention_characterization_revealed_preference} imply that $c_1$ is not a choice with competitive limited attention.

        \item [\textbf{CSLA, not COLA, not CCLA}]Let $c_2\colon \X\to X$ be the choice function on $X=\{w,x,y,z\}$ defined by
        $$
		w\underline{x}yz \; \; wx\underline{y} \; \; w\underline{x}z \; \; \underline{w}yz \; \;  \underline{x}yz \; \; w\underline{x} \; \; \underline{w}y \; \; \underline{w}z \;\;  x\underline{y} \; \; \underline{x}z \; \; \underline{y}z.
		$$
        Definition~\ref{DEF:revealed_salience} implies that $z \models_{c_2} w,x,y$ and $x \models_{c_2} w,y$.
        Thus, by Corollary~\ref{COR:salient_limited_attention_characterization_revealed_preference} $c_2$ is a choice with salient limited attention.
        Moreover, the reader can check that the pair $(\Gamma^{\,s},\rhd)$ such that $w\rhd y \rhd z\rhd x$ and $\Gamma^{\,s}(A)=\{v\in A\,\vert\, c_2(A)\rhd v\}\,\cup\,c(A)$ for any $A\in\X$ is an explanation by salient limited attention of $c_2$.
Definitions~\ref{DEF:revealed_shortlisting} and \ref{DEF:revealed_list} imply that $z R^{c_2} y R^{c_2} x R^{c_2} z$ and $x F_{c_2} y F_{c_2} x$.
By Corollaries~\ref{COR:optimal_limited_attention_characterization_revealed_preference} and \ref{COR:competitive_limited_attention_characterization_revealed_preference} we conclude that $c_2$ is neither a choice with optimal limited attention, nor a choice with competitive limited attention.
        
        \item [{CCLA, not COLA, not CSLA}]Let $c_3\colon \X\to X$ be the choice function on $X=\{w,x,y,z\}$ defined by
        $$
		wx\underline{y}z \; \;  wx\underline{y} \; \; \underline{w}xz \; \;  w\underline{y}z \; \;\underline{x}yz \; \; \underline{w}x \; \; w\underline{y} \; \;  w\underline{z}\;\;\underline{x}y \; \; \underline{x}z \; \; \underline{y}z. 
		$$
        The reader can check that the list $\mathbf{f}$ such that $y \,\mathbf{f}\, w \,\mathbf{f}\, x \,\mathbf{f}\, z$ list rationalizes $c_3$.
        Theorem~\ref{THM:list_rationalizable} implies that $c_3$ is a choice with competitive limited attention.
        Indeed, the pair $(\Gamma^{\,co},\mathbf{f})$ such that $\Gamma^{\,co}(A)=\{v\in A\,\vert\,c_3(A)\,\mathbf{f}\, v\} \,\cup\,c(A)$ for any $A\in\X$ is an explanation by competitive limited attention of $c_3$.
       Definition~\ref{DEF:revealed_salience} yields  $w \models_{c_3} x$ and $x \models_{c_3} w$, which implies by Corollary~\ref{COR:salient_limited_attention_characterization_revealed_preference} that $c_3$ is not a choice with salient limited attention. 
       Moreover, Definition~\ref{DEF:revealed_shortlisting} implies that $w R^{c_3} x R^{c_3} y R^{c_3}w$.
       Thus, by Corollary~\ref{COR:optimal_limited_attention_characterization_revealed_preference} $c_3$ is not a choice with optimal limited attention.   
       \end{description}
       
   Thus, any subclass is independent of the others.
  Moreover, the intersection between two specifications is independent of the third one, and it may generate attention filters that  simultaneously satisfy the requirements of each special case.
 The existence of the choice functions displayed below prove these facts.

       \begin{description}
        
        \item [\textbf{COLA, CSLA, not CCLA}]Let $c_4\colon \X\to X$ be the choice function on $X=\{w,x,y,z\}$ defined by
$$
		w\underline{x}yz \; \; \underline{w}xy \; \;w\underline{x}z \; \; w\underline{y}z \; \; \underline{x}yz \; \;  \underline{w}x \; \;w\underline{y} \; \; w\underline{z}\;\;\underline{x}y \; \; \underline{x}z \; \;  \underline{y}z.
		$$
        The pair $(P,\rhd)$ such that $xPy, zPw,$ and $y\rhd w\rhd x\rhd z,$ describes $c_4$.
        Thus, $c_4$ is a choice with optimal limited attention.
        Moreover, since $\vDash_{c_4}$ is defined by $z \models_{c_4} w,x,y,$ and $x \models_{c_4} w,y,$ it is asymmetric, and $c_4$ is a choice with salient limited attention.
        Indeed, the pair $(\Gamma,\rhd)$ such that $\Gamma(A)=\max(A,P)$ for any $A\in\X$ is an explanation by optimal limited attention of $c_4$.
       Moreover, $\Gamma$ is also a salient attention filter.
        To see this, note that in each of the menus $wxyz$, $wxy$, $wxz$, $wyz$, and $xyz,$ the most salient items, i.e. the worst alternative, and the best one among those considered, are respectively $xz, wy, xz, yz,$ and $xz.$
        According to $\Gamma$, the removal of any alternative distinct from them does not affect the DM's consideration set.
        For instance, $\Gamma(wxyz\setminus w)=\Gamma(xyz)=xz= \Gamma(wxyz)\setminus w$ holds.
        Finally, since $w F_{c_4} \,x$ and $x F_{c_4} w$, $F_{c_4}$ is not asymmetric, and $c_4$ is not a choice with optimal limited attention.

       \item [\textbf{COLA, CCLA, not CSLA}]Let $c_5\colon \X\to X$ be the choice function on $X=\{v,w,x,y,z\}$ defined by
        $$
        v\underline{w}xyz \;\; v\underline{w}xy \;\;  vwx\underline{z} \;\;  \underline{v}wyz \;\; v\underline{x}yz \;\; \underline{w}xyz \;\;  v\underline{w}x \;\;  \underline{v}wy \;\; vw\underline{z} \;\; v\underline{x}y \;\; vx\underline{z} \;\;  \underline{v}yz\;\;      
        $$
        $$
       \underline{w}xy \;\; wx\underline{z} \;\;  \underline{w}yz \;\; \underline{x}yz \;\;\underline{v}w\;\; v\underline{x} \;\; \underline{v}y \;\; v\underline{z} \;\; \underline{w}x \;\; \underline{w}y \;\; w\underline{z} \;\; \underline{x}y \;\; x\underline{z} \;\;  \underline{y}z.\;\;    
        $$
       The pair $(P,\rhd)$ such that   $x P v$, $y P z$, and $z\rhd v \rhd w \rhd x \rhd y$ describes $c_5$.
       Thus, $c_5$ is a choice with optimal limited attention.
      Moreover, the list $w\, \mathbf{f} \,v\,\mathbf{f} x\, \mathbf{f}\, z \,\mathbf{f}\, y$ list rationalizes $c_5$.
      Thus, $c_5$ is a choice with competitive limited attention.
      It is worth noting that the pair $(\Gamma^{\,opt},\rhd)$ such that $\Gamma^{\,opt}(A)=\max(A,P)$ for any $A\in\X$ is an explanation by optimal limited attention of $c_5$.
      However, fixed $\rhd$, $\Gamma^{\,opt}$ is not a competitive attention filter.
      To see why, note that $z\not\in\Gamma^{\,opt}(vwxyz)=wxy$, $w=\max(\Gamma^{\,opt}(vwxy),\rhd),$ but also $z\in\Gamma^{\,opt}(wz)=wz.$  
      On the other hand, the pair $(\Gamma^{\,co},\mathbf{f})$ such that $\Gamma^{\,co}(A)=\{t\in A\,\vert\, c(A) \,\mathbf{f}\,t\}\,\cup\,c(A)$ for any $A\in\X$ is a competitive attention filter, but it is not an optimal attention filter.
      To see this, note that $\Gamma^{\,co}(vwxyz)=vwxyz,$ but $\Gamma^{\,co}(vwxz)=z,$ which violates condition (ii)(c) of Definition \ref{DEF:rational_limited_consideration}. 
      Finally, observe that $x \models_{c_5} y$ and $y \models_{c_5} x$, which implies that $c_5$ is not a choice with salient limited attention.

        \item [\textbf{CSLA, CCLA, not COLA}] Let $c_6\colon \X\to X$ be the choice function on $X=\{w,x,y,z\}$ defined by
$$
		w\underline{x}yz \; \; wx\underline{y} \; \; w\underline{x}z \; \; w\underline{y}z \; \; \underline{x}yz \; \;   \underline{w}x \; \; w\underline{y} \; \; w\underline{z} \; \   \underline{x}y \; \; \underline{x}z \; \; \underline{y}z  .
		$$
        
       Since $\models_{c_6}$ is defined by $z \models_{c_6} w,x,y$ and $w \models_{c_6} x,y$, it is asymmetric, and $c_6$ is a choice with salient limited attention.
        Moreover, any list $\mathbf{f}$ such that $y \,\mathbf{f}\,w,x,$  and $ x\, \mathbf{f} \,w,z$ list rationalizes $c_6$.
        Thus, $c$ is a choice with competitive limited attention.
                Note that the pair $(\Gamma,\mathbf{f})$ such that $y\,\mathbf{f}\,x\,\mathbf{f} \,w\,\mathbf{f}\,z$, and $\Gamma=\{v\in A\,\vert\, c(A)\,\mathbf{f} \,v\}\,\cup\,c(A)$ for any $A\in\X$ is an explanation by competitive limited attention of $c_6$.
                 $\Gamma$ is also a salient attention filter.
                Indeed, note that in the menus $wxyz$, $wxy$, $wxz$, $wyz$, and $xyz,$ the most salient items with respect to $\mathbf{f}$  are respectively $xz, wy, xz, yz,$ and $xz.$
                According to $\Gamma$, the removal of any alternative distinct from them does not affect the DM's consideration.
                For instance, $\Gamma(wxyz\setminus w)=xz=\Gamma(wxyz)\setminus w.$
                Finally, note that $R^{c_6}$ is not acyclic, since $w R^{c_6} x, xR^{c_6} y$, and $yR^{c_6} w.$ 
                Thus, $c_6$ is not a choice with optimal limited attention.
         \end{description}
   Eventually, we present a choice function justified by any of the three specifications.     
        \begin{description}
                   \item [\textbf{COLA, CSLA, CCLA, not CSSLA}]Let $c_7\colon \X\to X$ be the choice function on $X=\{x,y,z\}$ defined by
         $$
         \underline{x}yz \;\; x\underline{y} \;\; \underline{x}z \;\; y\underline{z}.
         $$
         The pair $(P,\rhd)$ such that $z P y$, and $y\rhd x\rhd z$ describes $c_7$.
         Thus, $c_7$ is a choice with optimal limited attention, and the pair $(\Gamma,\rhd)$ such that $\Gamma(A)=\max(A,P)$ for any $A\in\X$ is an explanation by optimal limited attention of $c_7$.
         Moreover, $\vDash_{c_7}$ is asymmetric, since it is defined by $z\vDash_{c_7} x,y$, and $c_7$ is a choice with salient limited attention.
         Note also that any list such that $x\,\mathbf{f}\,y,z$ list rationalizes $c_7$, and $c_7$ is a choice with competitive limited attention.
         The pair $(\Gamma,\mathbf{f})$ such that $x\,\mathbf{f} \,y\,\mathbf{f} z$ is an explanation by competitive limited attention of $c_7$.
        $\Gamma$ is also a salient attention filter, and a competitive attention filter.
         Too see why, first note that in $xyz$ the most salient items with respect to $\mathbf{f}$ are $xz$, and $\Gamma(xyz\setminus y)=xz=\Gamma(xyz)\setminus y.$
         Second, observe that $x=\max(xyz,\mathbf{f}),$ $z=\max(\Gamma(xyz\setminus x),\mathbf{f}),$ $x\in\Gamma(xyz),$ and $x\in\Gamma(xz).$
        Finally, since $(xy,xyz)$ is a switch, $c_7$ is not rationalizable, and it is not a choice with selective salient limited attention.
        \end{description}

The analysis of $c_7$ shows two remarkable facts.
First, there is an optimal attention filter that, paired with some DM's preference, is also a salient attention filter, a competition filter, and justifies $c_7$.
Second, rationalizable choice behavior, or, alternatively, selective salient limited attention, is a proper subset of the intersection between optimal, salient, and competitive limited attention.
The main findings of this subsection are summarized in the following diagram.}

\vspace{1.5cm}
\begin{figure}[H]	
\begin{pspicture}(-5,-5)(5,6)
  \pscircle[ gradangle=45, gradbegin=blue!20, gradend=white, 
            linecolor=blue, linewidth=1pt](0.5,0){3.5} 

  \pscircle[ gradangle=45, gradbegin=red!20, gradend=white, 
            linecolor=red, linewidth=1pt](3.5,0){3.5}  

  \pscircle[ gradangle=45, gradbegin=green!20, gradend=white, 
            linecolor=green, linewidth=1pt](2,3){3.5} 

 \pscircle[linewidth=1pt, linecolor=magenta](2.03,1.2){1}
  
\pscircle[linewidth=1pt, linecolor=black](2.03,1.2){6}




 \rput(2,6.8){CLA}
  \rput(6,-0.3){CCLA}
  \rput(-2,-0.3){CSLA}
  \rput(2,5){COLA}
\rput(2.05,1.2){CSSLA}
\rput(2,4){$c_1$}
\rput(2,2.5){$c_7$}
\rput(0,2){$c_4$}
\rput(4,2){$c_5$}
\rput(2,-1.5){$c_6$}
\rput(-0.7,-1.5){$c_2$}
\rput(4.7,-1.5){$c_3$}
\end{pspicture}
\caption{The area delimited the black circle represents all the choices with limited attention, whereas the green, blue, red, and purple circles contain respectively all the choices with optimal, salient, competitive, and selective salient limited attention.
The choice functions $c_1, c_2, c_3, c_4, c_5, c_6, c_7$ are inserted in the intersections between specifications, following the indications of this subsection.}
\end{figure}

\subsection{Competitive limited attention and path independent consideration}\label{SUBSECT:path_independent_consideration}

Competitive attention filters recall \textit{competition filters}, i.e., choice correspondences requiring that any item selected in some menu must be selected in any subset of it.
\cite{LlerasMasatliogluNakajimaOzbay2017} introduce \textit{overwhelming choices}, which are choice functions whose selection from each menu is determined by the sequential application of a competition filter and some strict linear order.
The authors show that the class of overwhelming choices is equivalent to the model described by \cite{CherepanovFeddersenSandroni2013}, which is independent of choices with limited attention.
Moreover, in a separate work they analyze choice correspondences that satisfy \textit{path independence}, and the associated choice behavior.

\begin{definition}[\citealt{LlerasMasatliogluNakajimaOzbay2021}]
A choice correspondence $\Gamma^{\,PI}\colon\X\to \X$ satisfies \textit{Path Independence} if $\Gamma^{\,PI}(S\cup T)=\Gamma^{\,PI}(\Gamma^{\,PI}(S)\cup T)$ holds for any $S,T\in\X.$
A choice function $c\colon\X\to X$ is a $\pi$-\textit{LC model} if there is a {strict} linear order $\rhd$ and a choice correspondence $\Gamma^{\,PI}\colon\X\to \X$ satisfying Path Independence such that $c(A)=\max(\Gamma^{\,PI}(A),\rhd)$ for any $A\in\X.$
We say that $(\Gamma^{\,PI},\rhd)$ is an \textit{explanation by path independent limited consideration of} $c$. 
\end{definition}

A $\pi$-\textit{LC model} reports the picks of a DM whose consideration cannot be manipulated by the way in which menus are presented to her.
Path independent choice correspondences are characterized by the property of  competition filters, and that of attention filters.
Thus, any choice correspondence satisfying Path Independence is a competition filter, and the family of $\pi$-\textit{LC models} is the intersection between overwhelming choices and choices with limited attention.

The consideration reproduced by competitive attention filters is different from that described by competition filters and path independent choice correspondences.
Indeed, a competitive attention filter imposes on consideration sets some constraints that depend on \textit{pairs} of alternatives, and their ranking in the DM's preference.
Instead, a competition filter forces \textit{an} alternative that is observed in some menu to be considered in any subset of it.
A path independent choice correspondence postulates the coherence of the DM's consideration when she faces some menu that is the union of other menus already presented to her.
Due to {these} differences, choices with competitive limited attention and $\pi$-LC models are non-nested subclasses of limited attention. 
To prove this, 
we need some preliminary notation.
Given a choice function $c$, let $P^{\,PI}$ be the binary relation defined by $xP^{\,PI} y$ if there are $A,B\in\X$ such that $x,y\in A\subsetneq B,$ $x=c(A),$ and $c(B)\neq c(B\setminus y).$
\citet[Theorem 2 and Lemma A1]{LlerasMasatliogluNakajimaOzbay2021} prove that a choice function is a $\pi$-LC model if and only if $P^{\,PI}$ is asymmetric and acyclic.  

Consider the choice function $c_3,$ presented in Subsection~\ref{SECT:relationships_subclasses}\ref{SUBSECT:independence_of_spefication}.
The definition of $P^{\,PI}$ implies that $wP^{\,PI} x$, $x P^{\,PI} y $, and $y P^{\,PI} w.$
Thus, $c$ is not a $\pi$-LC model.
Moreover, $c_3$ is a choice with competitive limited consideration,  and  the pair $(\Gamma^{\,co},\rhd)$ such that $y\rhd w\rhd x\rhd z$, and  $\Gamma^{\,co}(A)=\{v\in A\,\vert\,c(A)\rhd v\}\,\cup\,c(A)$ for any $A\in\X$ 
is an explanation by competitive limited attention of $c_3$.
Note that $\Gamma^{\,co}$ does not satisfy Path Independence, since $wxy=\Gamma^{\,co}(wxy)\neq \Gamma^{\,co}(\Gamma^{\,co}(xy)\cup w)= \Gamma^{\,co}(x\cup w)=wx.$
Moreover, since $y\in\Gamma^{\,co}(wxy)$ and $y\not\in \Gamma^{\,co}(xy)$, $\Gamma^{\,co}$ is not a competition filter. 

Recall now that the choice function $c_4,$ defined in the previous subsection, is not a choice with competitive limited attention.
The relation $P^{\,PI}$, defined by $wP^{\,PI} x$, $x P^{\,PI} z,$ $y P^{\,PI} z,$ and $yP^{\,PI} w,$ is asymmetric and acyclic.
We conclude that $c_4$ is a $\pi$-LC model.
 Moreover, the pair $(\Gamma^{\,PI},\rhd)$ such that $y\rhd w \rhd x\rhd z$ and $\Gamma^{\,PI}$ is defined by
$$
		\Gamma^{\,PI}(wxyz)=xz \; \; \Gamma^{\,PI}(wxy)=wx \; \; \Gamma^{\,PI}(wxz)=xz
		 \; \; \Gamma^{\,PI}(wyz)=yz \; \;  \Gamma^{\,PI}(xyz)=xz $$
		 $$ \; \;  \Gamma^{\,PI}(wx)=wx \; \;\Gamma^{\,PI}(wy)=wy \; \; \Gamma^{\,PI}(wz)=z \; \ \Gamma^{\,PI}(xy)=x \; \;\Gamma^{\,PI}(xz)=xz \; \;\Gamma^{\,PI}(yz)=yz  $$

is an explanation by path independent limited consideration of $c_4$.  
However, $\Gamma^{\,PI}$ is not a competitive attention filter, since $y=\max(wxy,\rhd),$ $y\not\in\Gamma^{\,PI}(wxy)=wx,$ $w=\max(\Gamma^{\,PI}(wx),\rhd)$, and $y\in\Gamma^{\,PI}(wy)=wy.$

\section{Concluding remarks}\label{SECT:Concluding_remarks} 
In this work we show that many models of choices can be also interpreted through a different narrative, in which the DM's attention is limited, and it satisfies some behavioral properties.
To this end, we propose some subclasses of limited attention that explain only a portion of the observed choices justified by the general approach of \cite{MasatliogluNakajimaOzbay2012}, and we single out the characteristics of the associated attention filters. 
Moreover, for each special case, we partially identify the DM's unobserved preference and consideration sets from observed violations of the contraction consistency axiom.
We compare these findings with the information gathered by assuming that the DM behaves according to the examined model of choice.
Specifications are independent of each other, as well as their intersections.

This work contributes to a well-known debate in behavioral welfare economics concerning the multiplicity of patterns that explain the same choice behavior.
If, as established in  Theorems~\ref{THM:rational_attention_equivalent_shortlisting}, \ref{THM:CSLA_equivalent_representation}, \ref{THM:CSSLA_alpha}, and \ref{THM:list_rationalizable}, a specification of limited attention and some  rationality/bounded rationality model are indistinguishable in terms of choice data, we might ask which method should be the basis to perform a revealed preference analysis.
The use of limited attention rather than the equivalent models brings some advantages.
To begin with, there is a growing empirical and experimental evidence that supports  the specifications discussed in Subsections~\ref{SECT:limited_attention}\ref{SUBSECT:optimal_limited_consideration}, \ref{SECT:limited_attention}\ref{SUBSECT:Salient_limited_attention}, and \ref{SECT:limited_attention}\ref{SUBSECT:Competitive_limited_attention}.
As indicated by the methodological study of \cite{ManziniMariotti2014b}, these \textit{auxiliary data} may drive the experimenter to adopt a \textit{model-based approach}, and to assume that the DM's attention is limited.
In this case, the recognition of special cases of choice with limited attention, and investigation of the DM's preference and consideration sets become essential.
Such need is particularly compelling for list rationality, whose representation does not  involve the existence of any preference.
If contextual information shows that the DM has considered only some items of the feasible menus, then competitive limited attention offers an interpretation of a list rational choice that discloses her taste, and the bounds of her consideration.
Moreover, special cases of limited attention provide simple explanations of choices functions, by means of only two primitives.
Instead, equivalent models may demand the estimation of more parameters.
{For instance, any explanation by salient limited attention of the choice function $c_2$, defined in Subsection~\ref{SECT:relationships_subclasses}\ref{SUBSECT:independence_of_spefication} on four alternatives, consists of a linear order and a choice correspondence.
However, any general-temptation representation of $c_2$ is grounded on three real-valued functions.
Moreover, any conspicuity based endogenous reference representation of $c_2$ relies on a linear order, and four functions, each one attached to some item.}\footnote{{More precisely, $c_2$ has a general-temptation representation $(u,v,W)$ that must satisfy the inequalities $u(z)<u(w),u(x),u(y),$ and $u(x)<u(w),u(y).$
Moreover, $c_2$ admits each conspicuity based endogenous reference representation $(\gg,\{U_w,U_x,U_y,U_z\})$ respecting the conditions  $z\gg w,x,y,$ and $x\gg w,y$, $U_{z}(x)>U_{z}(w)>U_{z}(y)>U_{z}(z)$, $U_{x}(y)>U_{x}(x)>U_{x}(w),$ and either $U_{w}(w)>U_{w}(y)$ or $U_{y}(w)>U_{y}(y)$.}}
Beyond the dualism between alternative settings, there is also a synergy that results from combining them.
Indeed, limited attention corroborates the welfare conclusions about the DM's taste obtained from equivalent models.
Corollaries~\ref{COR:Optimal_limited_attention_shortlisting_same_preference} and~\ref{COR:list_competitive_limited_attention_same_preference} show that any preference retrieved from shortlisting and list rationality is consistent respectively with optimal and competitive limited attention.
Moreover, Lemma~\ref{LEM:common_preference_among_models} identifies some preference that is adopted by the DM, regardless of the method applied by the experimenter. 

We believe that our analysis may be extended in two directions.
 First, it would be useful to find limited attention representations of some stochastic choice behaviors.
A starting point may  be the stochastic extension of limited attention proposed by \cite{Cattaneoetal2020}, and  the general random behavioral model introduced by \cite{Aguiaretal2023}, called \textit{B-rule}, in which the DM's stochastic choice is explained by a pair of probability distributions that describe her randomizations over consideration sets and preferences.
The \textit{B-rule} has no empirical content, and the authors impose further conditions to retrieve testable patterns of limited attention.
Second, more advances in the elicitation of consideration sets from choice data are needed.
In the deterministic case this problem is structural, and full identification is not allowed.
However, the experimenter, once he selects a suitable specification, may estimate \textit{boundaries} to the DM's attention, by pinning down  respectively the smallest and largest attention filters that fit data.

\clearpage

\newpage

\section*{Appendix: proofs}

{\noindent{\underline{\it Proof of Remark 1}.}
 Given $X=\{x,y,z\}$, let $\Gamma\colon\X\to\X$ be the choice correspondence on $X$ defined by $\Gamma(A)=A$ for any $A\in\X\,\setminus \{x,y,z\}$, and $\Gamma(xyz)=yz.$
Conditions (ii)(a) and (ii)(c) hold for $\Gamma$, but (ii)(b) does not, since $x\in\Gamma(xy),$ $x\in\Gamma(xz),$ and $x\not\in\Gamma(xyz)$.
Consider now the choice correspondence $\Gamma^{\prime}\colon\X\to\X$  on $X$, defined by $\Gamma^{\prime}(A)=A$ for any $A\in\X\,\setminus \{x,y\}$, and $\Gamma^{\prime}(xy)=x,$ and note that (ii)(a) and (ii)(b) are true, but (ii)(c) is not satisfied, since $y\not\in\Gamma^{\prime}(xy),$ and $y\in \Gamma^{\prime}(xyz)$. 
Finally, given a set $X^{\prime}=\{w,x,y,z\}$, let $\Gamma^{\prime\prime}\colon\X^{\prime}\to\X^{\prime}$ be the choice correspondence on $X^{\prime}$ defined  by $\Gamma^{\prime\prime}(wxyz)=\Gamma^{\prime\prime}(wxy)=\Gamma^{\prime\prime}(wxz)=\Gamma^{\prime\prime}(wx)=\Gamma^{\prime\prime}(wz)=w,$
$\Gamma^{\prime\prime}(wyz)=\Gamma^{\prime\prime}(wy)=wy,$
$\Gamma^{\prime\prime}(xyz)=\Gamma^{\prime\prime}(xy)=\Gamma^{\prime\prime}(xz)=x,$ and $\Gamma^{\prime\prime}(yz)=y$.
$\Gamma^{\prime\prime}$ satisfies conditions (ii)(b) and (ii)(c), but (ii)(a) does not hold, since $y\not\in\Gamma^{\prime\prime}(wxy),$ and $\Gamma^{\prime\prime}(wxy)=w\neq wy= \Gamma^{\prime\prime}(wx)=wy$.\qed }

\medskip

\noindent{\underline{\it Proof of Theorem \ref{THM:rational_attention_equivalent_shortlisting}.}
We need three lemmas.
The first one formalizes a result mentioned in Section \ref{SECT:relationships_subclasses}, after Definition \ref{DEF:revealed_shortlisting}, in which the binary relation $R^c$ has been presented.

\begin{lemma}[\citealp{Yildiz2016}]\label{LEMMA:yldiz_shortlist_char}
	A choice $c$ is shortlisting if and only if $R^c$ is asymmetric and acyclic.
\end{lemma}

\begin{lemma}\label{LEMMA:limited_attention_holds_for_optimal_limited_attention}
	Any optimal attention filter is an attention filter. 
\end{lemma}

\begin{proof}
	Straightforward.
\end{proof}

\begin{lemma}\label{LEMMA:alpha_distance_1_implies_alpha_pairs}
	If a choice correspondence $\Gamma\colon\X\to\X$ satisfies property (ii)(c) of Definition~\ref{DEF:rational_limited_consideration}, then for any $x,y\in X$, and $B\in\X$,  $y \in \Gamma^{}(B \cup x)$ implies $y \in \Gamma^{}(xy).$ 
\end{lemma}

\begin{proof}
	Assume $y\in B,$ otherwise the statement is vacuously true.
Let $\vert (B\cup x)\setminus (xy)\vert=n$, and $ (B\cup x)\setminus (xy)=\{x_1,x_2,\cdots, x_n\}.$
By property (ii)(c) of Definition~\ref{DEF:rational_limited_consideration} we obtain that $y\in\Gamma((B\cup x)\setminus x_1).$
Apply again property (ii)(c) to conclude that $y\in \Gamma(((B\cup x)\setminus x_1)\setminus x_2).$
Since $X$ is finite, $n$ is finite, and we can use iteratively the same argument until we obtain that $y\in \Gamma((\cdots(B\cup x)\setminus x_1)\setminus x_2)\cdots)\setminus x_n)=\Gamma(xy).$ 	
\end{proof}

	\textit{\textbf{Only if part}}.
	Suppose that $c\colon\X\to X$ is a shortlisting.
	Thus, there is a pair $(>,\rhd)$ that describes $c$.
    Let $\Gamma^{\,opt}(A)=\max(A,>)$ for any $A\in\X$.\footnote{ Note that in the proof of \citet[Proposition 2]{Yildiz2016} the {strict} partial order $>$ is defined by using  a {strict} linear order $\rhd$ that extends $R^c$.
    Thus, $\Gamma^{\,opt}$ depends on $\rhd$.}
    We want to show that $(\Gamma^{\,opt},\rhd)$ is an explanation by optimal limited attention of $c$.
    Clearly $c(A)=\max(\Gamma^{\,opt}(A),\rhd)$ for any $A\in\X$.
    
    To prove that condition (ii)(a) of Definition \ref{DEF:rational_limited_consideration} holds, suppose that $x \notin  \Gamma^{\,\,opt}(B)$ for some $B\in\X$.
    The definition of $\Gamma^{\,\,opt}$ yields $x \notin \max(B,>)$.
    Thus, we obtain that $\Gamma^{\,\,opt}(B\setminus x)=\max(B\setminus x,>)=\max(B,>)\setminus x=\max(B,>)=\Gamma^{\,opt}(B)$.
    The proof that conditions  (ii)(b) and (ii)(c) of Definition \ref{DEF:rational_limited_consideration} are true is straightforward.

\textit{\textbf{If part}}.
  Assume that $c\colon\X\to X$ is a choice with optimal limited attention.
   Thus, there is a pair $(\Gamma^{\,opt},\rhd)$ that is an explanation by optimal limited attention of $c$.
     By Lemma~\ref{LEMMA:yldiz_shortlist_char}, to show that $c$ is a shortlisting, it is enough to prove that $R^c$ is asymmetric and acyclic, which, in turn, is implied by the fact that $R^c \sbs \rhd$.
    Thus, it is enough to show that $R^c \sbs \rhd$.
    Assume toward a contradiction that $R^c \not\sbs \rhd$.
   Thus, there are $x,y\in X$ such that  $\lnot(x \rhd y)$, hence $y \rhd x$, and $x R^c y$.
   
   If condition (i) of Definition~\ref{DEF:revealed_shortlisting}  holds, then there is $S\in\X$ such that $x=c(S \cup y)$ and $x \neq c(S)$.
    Since $y \rhd x$, we obtain $y \notin \Gamma^{\,opt}(S \cup y),$ and so by Lemma~\ref{LEMMA:limited_attention_holds_for_optimal_limited_attention} $\Gamma^{\,opt}(S \cup y)=\Gamma^{\,opt}(S)$, which yields $c(S \cup y)=c(S)$, a contradiction.
    
    If condition (ii) of  Definition~\ref{DEF:revealed_shortlisting} holds, then there is $S\in \X$ such that $y=c(S\cup x)$, and $x=c(xy)$.
    Definition~\ref{DEF:rational_limited_consideration} implies that $y\in \Gamma^{\,opt}(S \cup x)$ and $x=\max(\Gamma^{\,opt}(xy),\rhd)$.
    Moreover, condition (ii)(c) of Definition~\ref{DEF:rational_limited_consideration} and Lemma~\ref{LEMMA:alpha_distance_1_implies_alpha_pairs} imply that $y \in \Gamma^{\,opt}(xy)$, which, since $y \rhd x$, contradicts that $x=\max(\Gamma^{\,opt}(xy),\rhd)$.

    If condition (iii) of Definition~\ref{DEF:revealed_shortlisting} is true, then there is $S\in\X$ such that $y \neq c(S \cup x)$, $y=c(xy)=c(S)$.
    
    Definition~\ref{DEF:rational_limited_consideration} implies  that $y=\max(\Gamma^{\,opt}(xy),\rhd)=\max(\Gamma^{\,opt}(S),\rhd)$.
    By condition (ii)(b) of Definition~\ref{DEF:rational_limited_consideration} we have that $y \in \Gamma^{\,opt}(S \cup x)$.
    Since $y \neq c(S \cup x)$, there should be a $z \neq y$ such that $z=c(S\cup x)$, $z \in \Gamma^{\,opt}(S \cup x)$, and $z \rhd y$.
    If $z=x$, then $y \rhd x$ and $x \rhd y$, which is impossible. 
    If $z\neq x$, since $y=\max(\Gamma^{\,opt}(S),\rhd)$, we conclude that $z \notin \Gamma^{\,opt}(S)$, which 
     contradicts condition (ii)(c) of Definition~\ref{DEF:rational_limited_consideration}.}

\medskip
\noindent{\underline{\it Proof of Lemma \ref{LEMMA:necessary_conditions_of_RLC}.}
By Theorem~\ref{THM:rational_attention_equivalent_shortlisting} $c$ is a shortlisting, and there is a pair $(>,\rhd)$ that describes $c$.
Since $x\neq c(A)\neq c(A\setminus x)$, by Definition~\ref{DEF:shortlisting} two cases are possible:
\begin{enumerate}[\rm(i)]
	\item $c(A)\rhd c(A\setminus x)$, and $c(A)\not\in \max(A\setminus x,>)$, or
	\item $c(A\setminus x)\rhd c(A)$, and $c(A\setminus x)\not\in\max(A,>)$.
\end{enumerate}
If (i) holds, there is $y\in A\setminus x$ such that $y>c(A)$, which implies that $c(A)\not\in\max(A,>)$, which is impossible.
Thus (ii) must hold.
Note that, since $c(A\setminus x)\not\in\max(A,>)$, and $c(A\setminus x)\in \max(A\setminus x,>)$, we conclude that $x> c(A\setminus x)$.
Moreover, $x\in\max(A,>)$, otherwise there should be a $y\in A\setminus x$ such that $y>x>c(A\setminus x)$, implying, since $>$ is transitive, that $y>c(A\setminus x)$ and that $c(A\setminus x)\not\in\max(A\setminus x,>),$ which is impossible.

Let $(\Gamma^{\,opt},\rhd)$ be some explanation by  optima} limited attention of $c$.
Since $x\neq c(A)\neq c(A\setminus x)$, Definition~\ref{DEF:rational_limited_consideration} and  Lemma~\ref{LEMMA:limited_attention_holds_for_optimal_limited_attention} yield $\Gamma^{\,opt}(A)\setminus x\neq \Gamma^{\,opt}(A\setminus x)$, $x\in\Gamma^{\,opt}(A)$ and $c(A)\rhd x$.
Note also that, since $c(A)\in\Gamma^{\,opt}(A)$, by condition (ii)(c) of Definition~\ref{DEF:rational_limited_consideration} and Lemma~\ref{LEMMA:alpha_distance_1_implies_alpha_pairs} we conclude that $c(A)\in\Gamma^{\,opt}(c(A)x).$ 
Moreover, two cases are possible:
\begin{enumerate}
	\item[(iii)] $c(A)\rhd c(A\setminus x)$, and $c(A)\not\in \Gamma^{\,opt}(A\setminus x)$, or
	\item[(iv)] $c(A\setminus x)\rhd c(A)$, and $c(A\setminus x)\not\in \Gamma^{\,opt}(A)$.
\end{enumerate}

Condition (iii) implies that $\Gamma(A)\setminus x\not\subseteq \Gamma (A\setminus x), $ which contradicts condition (ii)(c) of Definition~\ref{DEF:rational_limited_consideration}.
Thus, (iv) holds, and, since $c(A\setminus x)\not\in \Gamma^{\,opt}(A)$ and $c(A\setminus x)\in  \Gamma^{\,opt}(A\setminus x)$, by condition (ii)(b) of Definition~\ref{DEF:rational_limited_consideration} we must have that $c(A\setminus x)\not\in\Gamma^{\,opt}(c(A\setminus x)x).$ 
\qed

\medskip

\medskip
\noindent{\underline{\it Proof of Lemma \ref{LEM:partial_order_equals_optimal_attention_filter}.}
We need some preliminary definitions, and a  result.

\begin{definition}[\citealt{CantoneGiarlottaGrecoWatson2016}]
	A choice correspondence $\Gamma \colon \X \to \X$ is
\textsl{quasi-transitively rationalizable} if there exists a {strict} partial order $>$ on $X$ such that $\Gamma(A)= \max(A,>)$ for any $A \in \X$.
\end{definition}
 
\begin{definition}
Let $\Gamma\colon\X\to\X$ be a choice correspondence on $X$.
$\Gamma$ \textit{satisfies Axiom$\:\alpha\,$} if  for all $A,B\in \X$ and $x \in X$, if $x \in A \subseteq B$ and $x\in \Gamma(B)$, then $x\in \Gamma(A)$.\footnote{This is the natural generalization to choice correspondences of Contraction Consistency.}
$\Gamma$ \textit{satisfies Axiom$\:\gamma\,$} if 
  for all $A,B\in \X$ and $x\in A\cap B$,  if $x\in \Gamma(A)\cap \Gamma(B)$, then $x\in\Gamma(A\cup B)$. 
Finally, $\Gamma$ \textit{satisfies	Axiom$\:\delta\,$}
  for all $A,B\in \X$ and  distinct $x,y\in X$, if $A\subseteq B$ and $x,y\in \Gamma(A)$, then $x\neq\Gamma(B)\neq y$. 
\end{definition}

\begin{theorem}[\citealt{Sen1986}]\label{THM:quasi_transitively_rationalizability_characterization}
	A choice correspondence $\Gamma\colon\X\to\X$ on $X$ is quasi-transitively rationalizable if and only if it satisfies Axioms$\:\alpha\,,$ $\gamma,$ and $\delta.$
\end{theorem}

We are now ready to prove Lemma \ref{LEM:partial_order_equals_optimal_attention_filter}.
In the proof of Theorem~\ref{THM:rational_attention_equivalent_shortlisting} we also show that (i) implies (ii).
Given Theorem~\ref{THM:quasi_transitively_rationalizability_characterization}, to show that (ii) implies (i) we are only left to show that any optimal attention filter $\Gamma^{\,opt}\colon\X\to\X$ on $X$ satisfies Axioms$\:\alpha\,$ $\gamma,$ and $\delta.$
To see that $\Gamma$ satisfies Axiom$\:\alpha,$ assume that $x\in \Gamma^{\,opt}(B),$ for some $B\in\X,$ and $x\in B$.
Let $A\subseteq B$ some subset of $B$ such that $x\in A$.
 Let $\vert B\setminus A\vert=n,$ and $B\setminus A=\{y_1,y_2,\cdots,y_n\}.$
 Condition (ii)(c) of Definition~\ref{DEF:rational_limited_consideration} implies that $x\in \Gamma^{\,opt}(B\setminus y_1).$
 Apply again condition (ii)(c) to conclude that $x\in  \Gamma^{\,opt}((B\setminus y_1)\setminus y_2).$
 Since $X$ is finite, $n$ is finite, and we can use iteratively the same argument until we obtain that $x\in \Gamma^{\,opt}((\cdots(B\setminus y_1)\setminus y_2)\cdots)\setminus y_n)=\Gamma^{\,opt}(A).$

To prove that $\Gamma^{\,opt}$ satisfies Axiom$\:\gamma,$ suppose that there are $A,B\in\X$, and $x\in A\cap B$ such that $x\in \Gamma^{\,opt}(A)\cap\Gamma^{\,opt}(B)$.
Let $\vert A\setminus B\vert=n,$ $\vert B\setminus A\vert=m,$ $A\setminus B=\{y_1,\cdots,y_n\},$ and $B\setminus A=\{z_1,\cdots,z_m\}$.
Condition (ii)(c) of Definition~\ref{DEF:rational_limited_consideration} and Lemma~\ref{LEMMA:alpha_distance_1_implies_alpha_pairs} implies that $x\in\Gamma^{\,opt}(xy_i)$ for any $y_i\in A\setminus B,$ and $x\in\Gamma^{\,opt}(xz_i)$ for any $z_i\in B\setminus A.$
Since $\vert X\vert$ is finite, $m$ and $n$ are finite, and we can apply iteratively condition (ii)(b) of Definition~\ref{DEF:rational_limited_consideration} to conclude that $x\in\Gamma^{\,opt}(A\cup B).$

Finally, to show that $\Gamma^{\,opt}$ satisfies Axiom$\,\delta,$ assume that there are $A,B\in\X$, and distinct $x,y\in A$ such that $A\subseteq B,$ and $x,y\in\Gamma^{\,opt}(A).$
Let $\vert B\setminus A\vert=n,$  and $B\setminus A=\{z_1,z_2,\cdots,z_n\}.$
Now suppose toward a contradiction that $\Gamma^{\,opt}(B)=x.$
Condition (ii)(a) implies that $x=\Gamma^{\,opt}(B)=\Gamma^{\,opt}(B\setminus z_1).$
Apply again condition (ii)(a) to conclude that $x=\Gamma^{\,opt}(B\setminus z_1)=\Gamma^{\,opt}((B\setminus z_1)\setminus z_2).$
Since $n$ is finite, we can apply iteratively this argument until we get $x=\Gamma^{\,opt}((\cdots(B\setminus z_1)\setminus z_2)\cdots)\setminus z_n)=\Gamma^{\,opt}(A)=x,$  which is false.
We can apply the same argument when we assume that $\Gamma(B)=y,$ and we would obtain a contradiction again.
Thus, we conclude that $x\neq \Gamma(B)\neq y.$

\medskip

\noindent{\underline{\it Proof of Lemma \ref{LEM:existence_salience_attention_filter_attention_filter}.}
 Let $c\colon \X\to X$ be a choice with salient limited attention, and let $(\Gamma^{s},\rhd)$ be some explanation by salient limited attention of $c$.
Two cases are possible: (i) $\min(A,\rhd)\in\Gamma^{\,s}(A)$ for any $A\in \X$, or (ii) $\min(A,\rhd)\not\in\Gamma^{\,s}(A)$ for some $A\in\X$.
If (i) holds, then we know that, for any $B\in \X$ and $x\in X$, $x\not\in\Gamma^{\,s}(B)$ yields $x\neq \min(B,\rhd),\max(\Gamma^{\,s}(B),\rhd)$.
Definition~\ref{DEF:CSLA} implies that $\Gamma^{\,s}(B)=\Gamma^{\,s}(B)\setminus x=\Gamma^{s}(B\setminus x)$.
If (ii) holds, then let ${\Gamma^{\,s}}^{\prime}\colon\X\to \X$ be the choice correspondence defined by ${\Gamma^{\,s}}^{\prime}(A)=\Gamma^{s}(A)\,\cup \min(A,\rhd)$ for any $A\in\X$.
We show that $({\Gamma^{\,s}}^{\prime},\rhd)$ is an explanation by salient limited attention of $c$.
Consider some $B\in\X,$ and $x\in B$, and assume that $x\neq  \min(B,\rhd),\max({\Gamma^{\,s}}^{\prime}(B),\rhd)$.
The definition of ${\Gamma^{\,s}}^{\prime}$ yields ${\Gamma^{\,s}}^{\prime}(B)\setminus x=(\Gamma^{\,s}(B)\cup  \min(B,\rhd))\setminus x$ and $\Gamma^{\,s}(B\setminus x)\cup  \min(B,\rhd)={\Gamma^{\,s}}^{\prime}(B\setminus x)$.
Since $\Gamma^{\,s}$ is a salient attention filter we obtain that $(\Gamma^{\,s}(B)\cup  \min(B,\rhd))\setminus x=\Gamma^{\,s}(B\setminus x)\cup  \min(B,\rhd).$
We conclude that ${\Gamma^{\,s}}^{\prime}(B)\setminus x={\Gamma^{\,s}}^{\prime}(B\setminus x),$ and that ${\Gamma^{\,s}}^{\prime}$ is a salient attention filter.
Moreover, $c(B)=\max(\Gamma^{\,s}(B),\rhd)=\max(\Gamma^{\,s}(B)\cup\min(B,\rhd),\rhd)=\max({\Gamma^{\,s}}^{\prime}(B),\rhd)$ for any $B\in\X$.
Finally, since $\min(A,\rhd)\in{\Gamma^{\,s}}^{\prime}(A)$ for any $A\in \X$, we are back to case (i), and we can conclude that ${\Gamma^{\,s}}^{\prime}$ is an attention filter.

\medskip
\noindent\underline{\it Proof of Theorem~\ref{THM:CSLA_equivalent_representation}.}
The equivalence between (ii) and (iii) has been proved in \citet[Theorem~3]{GiarlottaPetraliaWatson2022b}.
This result also contains a third equivalent condition, i.e., the asymmetry of the binary relation $\vDash_c$, introduced in Definition~\ref{DEF:revealed_salience}.   
Note also that, according to \citet[Proposition~1]{RavidStevenson2021}, a choice function $c\colon\X\to X$  admits a general temptation representation if and only if the \textit{Axiom of Revealed Temptation (ART)} holds.
ART requires that for any $B\in\X$, there is $x\in B$ such that  any pair $(B^{\prime},B^{\prime\prime})$ with $x \in B^{\prime}\subsetneq B^{\prime\prime}\subseteq B$ is not a switch.
Thus, to complete the proof of Theorem~\ref{THM:CSLA_equivalent_representation} we only need to show that the asymmetry of $\vDash_c$ is equivalent to ART.

Let $c\colon\X\to X$ be a choice function.
We first show that when $\vDash_c$ is asymmetric, ART holds.
Assume that $\vDash_c$ is asymmetric.
By \citet[Lemma 4]{GiarlottaPetraliaWatson2022b}, we obtain that $\vDash_c$ is acyclic, hence a {strict} partial order.
By \cite{Szpilrajn1930}'s theorem,  there is a {strict} linear order $\rhd$ that extends $\vDash_c$.
We claim that, for any $B\in \X$, any pair  $(B^{\prime},B^{\prime\prime})$ such that $\max(B,\rhd)\in B^{\prime}\subsetneq B^{\prime\prime}\subseteq B$ is not a switch. 
Toward a contradiction, assume there are $B^{\prime}\subsetneq B^{\prime\prime}\subseteq B$ such that $\max(B,\rhd)\in B^{\prime}$ and $(B^{\prime}, B^{\prime\prime})$ is a switch. 
By Lemma~\ref{LEMMA:minimal_violations_of_alpha}, there are $y \in X$ and $C \in \X$ such that $B^{\prime}\subseteq C \setminus y \subsetneq C \subseteq B^{\prime\prime}$ and $(C\setminus y,C)$ is a minimal switch. 
Note that $y$ must be distinct from $\max(B,\rhd)$.
The definition of $\vDash_c$ implies that $y \vDash_c x$, and that $y\rhd x$, a contradiction. 

To prove that ART implies the asymmetry of $\vDash_c$, assume toward a contradiction that ART holds, and $\vDash_c$ is not asymmetric.
Thus, there are $x,y\in X$, and $D,E\in \X$ such that $x,y\in D\cap E$, $(D\setminus x,D)$ is a minimal switch, and $(E\setminus y,E)$ is a minimal switch.
Consider the menu $D\cup E$.
Axiom $\alpha$ does not hold for the collection $\{G \subseteq D\cup E \colon x\in G\}$, since $(E\setminus y, E)$ is a minimal switch.
Axiom\;$\alpha$ does not hold for the collection $\{G \subseteq D\cup E \colon y\in G\}$, since $(D\setminus x, D)$ is a minimal switch.
Moreover, for any $z\in D\setminus \{x,y\}$ Axiom\;$\alpha$ fails for the collection $\{G \subseteq D\cup E \colon z\in G\}$, since $(D\setminus x, D)$ is a minimal switch.
Finally for any $z\in E\setminus \{x,y\}$ Axiom\;$\alpha$ fails for the collection $\{G \subseteq D\cup E \colon z\in G\}$, since $(E\setminus y, E)$ is a minimal switch.
We conclude that ART fails for $c$, a contradiction.
\qed

\medskip
\noindent\underline{\it Proof of Lemma~\ref{LEM:CSLA_partial_identification}.}
By Theorem~\ref{THM:CSLA_equivalent_representation} $c$ has a general-temptation representation $(u,v,W)$.  
Since $x\neq c(A)\neq c(A\setminus x)$, by Definition~\ref{DEF:temptation} we must have that $$W(u(c(A\setminus x)),v(c(A\setminus x))-\max_{y\in A\setminus x} v(y)))>W(u(c(A)),v(c(A))-\max_{y\in A\setminus x} v(y))),$$ and $$W(u(c(A)),v(c(A))-\max_{y\in A} v(y)))> W(u(c(A\setminus x)),v(c(A\setminus x))-\max_{y\in A} v(y))),$$
which implies that $v(x)=\max_{y\in A} v(y)$.
Moreover, four cases are possible: 
\begin{enumerate}[\rm(i)]
		\item   $u(c(A))\leq u(c(A\setminus x))$, and $v(c(A))\leq v(c(A\setminus x))$, or 
	\item $u(c(A))\geq u(c(A\setminus x))$, and $v(c(A))\geq v(c(A\setminus x))$, or
\item $u(c(A)) < u(c(A\setminus x))$, and $v(c(A))> v(c(A\setminus x))$, or
\item  $u(c(A))> u(c(A\setminus x))$, and $v(c(A)) < v(c(A\setminus x))$.
\end{enumerate}

If (i) or (ii)  holds, Definition~\ref{DEF:temptation} implies that $c(A)=c(A\setminus x)$, which is false. 
Thus, we conclude that either (iii) or (iv) hold.

Theorem~\ref{THM:CSLA_equivalent_representation} implies that $c$ admits a conspicuity based endogenous reference representation $(\gg,\{U_{z}\}_{z\in X})$.
Since $x\neq c(A\setminus x)\neq c(A)$, Definition~\ref{DEF:endogenous_reference_representation} yields $x=\max(A,\gg)\neq\max(A\setminus x,\gg)$, and $x\gg y$ for any $y\in A$, and that $U_{x}(c(A))>U_{x}(y)$ for any $y\in A$.

Finally let $(\Gamma^{\,s},\rhd)$ be some explanation by salient limited attention of $c$.
Assume that there $A\in\X$ and $x\in A$ such that  $(A\setminus{x},A)$ is a minimal switch, i.e. $c(A\setminus x)\neq c(A)\neq x$.
 Definition~\ref{DEF:CSLA} implies that $\Gamma^{\,s}(A\setminus x)\neq \Gamma^{\,s}(A)$, and, since $x\in A$ and $x\neq c(A)$, that $x\in\Gamma^{\,s}(A)$, $\Gamma^{\,s}(A\setminus x)\neq \Gamma^{\,s}(A)\setminus x$, and $x=\min(A,\rhd)$.
\medskip

{\noindent\underline{\it Proof of Lemma~\ref{LEM:common_preference_among_models}.}
We need some preliminary notions and results.
In the proof of \citet[Theorem 3]{GiarlottaPetraliaWatson2022b} the following fact is shown.

\begin{lemma}\label{LEM:salient_limited_attention_filter_extension}
Let $c\colon\X \to X$ and $\rhd$ be respectively a choice function and a {strict} linear order on $X$.
Moreover, let $\Gamma_{\rhd}\colon\X\to X$ be defined by $\Gamma_{\rhd}(A)=\{x\in A\,\vert\, c(A)\rhd x\}\,\cup\,c(A)$ for any $A\in\X.$
	If $\widetilde{P}\subseteq \rhd$, then $(\Gamma_{\rhd},\rhd)$ is an explanation by salient limited attention of $c$. 
\end{lemma}

\begin{lemma}\label{LEM:revealed_salience_does_not_hold_in_the_tail}
Let $c\colon \X\to X$ be a choice function on $X$.
Assume that the binary relation $\vDash_c$ is asymmetric and acyclic.
 For any {strict} linear order $\gg$ that extends $\vDash_c,$ we have that $\neg \left(x^{\gg}_{\vert X\vert-1}\vDash_c x^{\gg}_{\vert X\vert}\right)$ and $\neg \left(x^{\gg}_{\vert X\vert}\vDash_c x^{\gg}_{\vert X\vert-1}\right).$
 \end{lemma}
 
 \begin{proof}
 The fact that $\neg \left(x^{\gg}_{\vert X\vert}\vDash_c x^{\gg}_{\vert X\vert-1}\right)$ is obvious.
Thus, assume toward a contradiction that $\gg$ extends $\vDash_c,$ and $x^{\gg}_{\vert X\vert-1} \vDash_c x^{\gg}_{\vert X\vert}.$
 Definition~\ref{DEF:revealed_salience} implies that there is a menu $A\in\X$ such that $x^{\gg}_{\vert X\vert}\in A$, and $(A\setminus x^{\gg}_{\vert X\vert-1}, A)$ is a minimal switch.
 If $\vert A\vert>2,$ then it would exist some $y\in A,$ distinct from $x^{\gg}_{\vert X\vert-1}$ and $x^{\gg}_{\vert X\vert},$ such that $x^{\gg}_{\vert X\vert-1}\vDash_c y.$
 Since $\gg$ extends $\vDash_c,$ we should conclude that $x^{\gg}_{\vert X\vert-1}\neq \min(X\setminus \min(X,\rhd),\gg),$ which is false.
 Thus, $\vert A\vert =2,$ which implies that $(A\setminus x^{\gg}_{\vert X\vert-1}, A)$ is not a switch, a contradiction.  
 \end{proof}

\begin{definition}
	\label{DEF:inverse_salience_and_incomparability}
	Let $c\colon \X\to X$ be a choice function on $X$.
	Let $\gg$ be some {strict} linear order, if any, extending $\vDash_c.$
	Let ${\widetilde{P}}^{*}\supseteq \widetilde{P}$ be the relation defined, for any $x,y\in X,$ by $x{\widetilde{P}}^{*}y$ if and only if $x\widetilde{P}y$ or $x=x^{\gg}_{\vert X\vert}$ and $y=x^{\gg}_{\vert X\vert-1}.$
	Moreover, denote by  ${\widetilde{P}}^{**}\supseteq \widetilde{P}$  the relation defined, for any $x,y\in X,$ by $x{\widetilde{P}}^{**}y$ if and only if $x\tilde{P}y$ or $x=x^{\gg}_{\vert X\vert-1}$ and $y=x^{\gg}_{\vert X\vert}.$
	
\end{definition}

 \begin{lemma}\label{LEMMA:asymmetry_acyclicity_P*_P**}\label{LEM:inverse_revealed_salience_with_incomparable_items_is_asymmetric_acyclic}
 	If $\vDash_c$ is asymmetric and acyclic, then ${\widetilde{P}}^{*}$ and ${\widetilde{P}}^{**}$ are asymmetric and acyclic.
 	\end{lemma}
 	
 \begin{proof}
 Since $\vDash_c$ is asymmetric and  acyclic, by \cite{Szpilrajn1930}'s theorem there is some {strict} linear order $\gg$ extending ${\widetilde{P}}^{*}$.
 Thus, $x^{\gg}_{\vert X\vert-1}$ and $x^{\gg}_{\vert X\vert}$ exist, $\widetilde{P}\subsetneq {\widetilde{P}}^{*}$, and $\widetilde{P}\subsetneq {\widetilde{P}}^{**}$.
 The definition of $\widetilde{P}$ implies that $\widetilde{P}$ is asymmetric and acyclic.
 Moreover, Definition~\ref{DEF:inverse_salience_and_incomparability},
 Lemma~\ref{LEM:revealed_salience_does_not_hold_in_the_tail}, and the asymmetry of ${\widetilde{P}}$  imply that ${\widetilde{P}}^{*}$ and ${\widetilde{P}}^{**}$ are asymmetric.
 
 To show that ${\widetilde{P}}^{*}$ is acyclic, assume toward a contradiction that there are items $x_1,\cdots,x_n\in X$ such that $n\geq 3,$ and $x_1{\widetilde{P}}^{*} \cdots {\widetilde{P}}^{*} x_n\widetilde{P}^{*} x_1.$
 Without loss of generality, we can assume that $x_1=x^{\gg}_{\vert X\vert}$ and $x_2=x^{\gg}_{\vert X\vert-1}.$
 Otherwise, Definition~\ref{DEF:inverse_salience_and_incomparability} would imply that $x_1\widetilde{P} \cdots \widetilde{P} x_n\widetilde{P} x_1,$ which contradicts the acyclicity of $\tilde{P}.$
We obtain that $x_n\widetilde{P}^{*}x_1.$
The asymmetry of $\widetilde{P}^{*}$ implies that $x_n\neq x^{\gg}_{\vert X\vert-1}$.
Since $x_n\neq x^{\gg}_{\vert X\vert-1}$, Definition~\ref{DEF:inverse_salience_and_incomparability} yields $x_n\, \widetilde{P}\,x^{\gg}_{\vert X\vert}.$
The definition of $\widetilde{P}$ implies that $x^{\gg}_{\vert X\vert}\vDash_c x_n.$
Since $\gg$ extends $\vDash_c$, we conclude that $x^{\gg}_{\vert X\vert}\gg x_n,$ which is impossible.

 To show that ${\widetilde{P}}^{**}$ is acyclic, assume toward a contradiction that there are items $x_1,\cdots,x_n\in X$ such that $n\geq 3,$ and $x_1{\widetilde{P}}^{**} \cdots {\widetilde{P}}^{**} x_n\widetilde{P}^{**} x_1.$
 Without loss of generality, we can assume that $x_1=x^{\gg}_{\vert X\vert-1}$ and $x_2=x^{\gg}_{\vert X\vert}.$
 Otherwise, Definition~\ref{DEF:inverse_salience_and_incomparability} would imply that $x_1\widetilde{P} \cdots \widetilde{P} x_n\widetilde{P} x_1,$ which contradicts the acyclicity of $\tilde{P}.$
 We obtain that $x_n\widetilde{P}^{**}x_1.$
The asymmetry of $\widetilde{P}^{**}$ implies that $x_n\neq x^{\gg}_{\vert X\vert}$.
Since $x_n\neq x^{\gg}_{\vert X\vert}$, Definition~\ref{DEF:inverse_salience_and_incomparability} yields $x_n\, \widetilde{P}\,x^{\gg}_{\vert X\vert-1}.$
The definition of $\widetilde{P}$ implies that $x^{\gg}_{\vert X\vert-1}\vDash_c x_n.$
Since $\gg$ extends $\vDash_c$, we conclude that $x^{\gg}_{\vert X\vert-1}\gg x_n,$ which is impossible, since $x_n\neq x^{\gg}_{\vert X\vert}$.  \end{proof}

We are ready to prove Lemma~\ref{LEM:common_preference_among_models}.
Let $c\colon \X\to X$ be a choice with salient limited attention.
We first show that $\rhd^{\tilde{P},\gg}$ exists for some {strict} linear order $\gg$ on $X$, and then we prove that $\rhd^{\tilde{P},\gg}$ satisfies conditions (i), (ii), and (iii).

By Theorem~\ref{THM:CSLA_equivalent_representation} $c$ admits a conspicuity based endogenous reference representation, and by \citet[Theorem~3]{GiarlottaPetraliaWatson2022b} and \citet[Lemma~4]{GiarlottaPetraliaWatson2022b} $\vDash_c$ is asymmetric and acyclic.
Thus, by \cite{Szpilrajn1930}'s theorem there is a {strict} linear order $\gg$ extending $\vDash_c$.
Moreover, Definition~\ref{DEF:inverse_salience_and_incomparability},  Lemma \ref{LEMMA:asymmetry_acyclicity_P*_P**}, and \cite{Szpilrajn1930}'s theorem imply that  $\rhd^{\widetilde{P},\gg}$  always exists.

To verify that $\rhd^{\widetilde{P},\gg}$ satisfies (i), note that \citet[Proposition 3]{RavidStevenson2021} show that, if ART holds, then, for any {strict} linear order $\rhd$ that extends $\widetilde{P},$ there is a general-temptation representation $(u,v,W)$ such that, for any $x,y\in X$, $u(x)>u(y)$ holds if and only if $x\rhd y$ is true.

Next we prove that $\rhd^{\widetilde{P},\gg}$ satisfies (ii).
For any $k\in X\setminus x^{\gg}_{\vert X\vert-1},$ let $>_k$ be the binary relation defined  by $x>_k y$ if there is a menu $A\in \X$ such that $k=\max(A,\gg),$ $y\in A,$ and $c(A)=x.$
\citet[Lemma 13]{GiarlottaPetraliaWatson2022b} show that, if $\vDash_c$ is asymmetric, then $>_k$ is a {strict} partial order, which, by \cite{Szpilrajn1930}'s theorem can be extended by some {strict} linear order $\rhd_k$ on $X$.
 Let $\{U_k\}_{k\in X\setminus x^{\gg}_{\vert X\vert-1}}$ be some family of real valued functions such that, for any $k\in X\setminus x^{\gg}_{\vert X\vert-1}$, $U_{k}(x)>U_{k}(y)$ holds if and only if $x \rhd_{k} y$ holds, with $\rhd_k$ being some {strict} linear order extending $>_k$.
 Since each $\rhd_k$ is a {strict} linear order, any $U_k$ is injective.
 Moreover, let $\{U_z\}_{z\in X}$ be defined by $\{U_k\}_{k\in X\setminus x^{\gg}_{\vert X\vert-1}}\cup \left\{U_{x^{\gg}_{\vert X\vert-1}}\right\},$ where $U_{x^{\gg}_{\vert X\vert-1}}$ is some real valued function such that $U_{x^{\gg}_{\vert X\vert-1}}(x)>U_{x^{\gg}_{\vert X\vert-1}}(y)$ holds if and only if $x\rhd^{\widetilde{P},\gg} y$ is true.
 Since $\rhd^{\widetilde{P},\gg}$ is a {strict} linear order, $U_{x^{\gg}_{\vert X\vert-1}}$ is injective.  
 We show now that $(\gg, \{U_z\}_{z\in X})$ is a conspicuity based endogenous reference representation of $c.$
For any menu $A\in \X$ distinct from $\left\{x^{\gg}_{\vert X\vert-1},x^{\gg}_{\vert X\vert}\right\} $ and $\{x^{\gg}_{\vert X\vert-1}\}$,
the definition of $\{U_k\}_{k\in X\setminus x^{\gg}_{\vert X\vert-1}}$ implies that $c(A)=\argmax_{x\in A}U_{\max(A,\gg)}(x)$.
Moreover, $c\left(x^{\gg}_{\vert X\vert-1}x^{\gg}_{\vert X\vert}\right)=\argmax_{x\in \left\{x^{\gg}_{\vert X\vert-1},x^{\gg}_{\vert X\vert}\right\}}U_{x^{\gg}_{\vert X\vert-1}}\left(x\right),$ and $c\left(x^{\gg}_{\vert X\vert-1}\right)=\argmax_{x\in \left\{x^{\gg}_{\vert X\vert-1}\right\}}U_{x^{\gg}_{\vert X\vert-1}}\left(x\right).$

Finally, to see that $\rhd^{\widetilde{P},\gg}$ satisfies (iii), 
note that by Lemma~\ref{LEM:salient_limited_attention_filter_extension} $\left(\Gamma_{\rhd^{\widetilde{P},\gg}},\rhd^{\widetilde{P},\gg}\right)$ is an explanation by salient limited attention of $c.$}

 \medskip
 
 \noindent\underline{\it Proof of Theorem~\ref{THM:CSSLA_alpha}.}
\textit{\textbf{Only if part}}. Assume $c\colon\X\to X$ satisfies Axiom\;$\alpha$.
According to \cite{Samuelson1938}, there is a {strict} linear order $\rhd$ on $X$ that such that $c(A)=\max(A,\rhd)$ for each $A\in\X$.
Let $x_{\,\min}=\min(X,\rhd)$ be the minimum element of $X$ with respect to $\rhd$.
Let $\Gamma^*\colon\X\to \X$ be a choice correspondence such that $\Gamma^*(A)=A\setminus x_{\,\min}$, for any $A\in\X\setminus \{x_{\,\min}\}$, and $\Gamma^*(x_{\,\min})=x_{\,\min}$.
We need to show that $\Gamma^*$ is a selective attention filter, and that $(\Gamma^*,\rhd)$ is an explanation by selective salient limited attention of $c$.\footnote{This result would have been valid even if we had defined $\Gamma^*(A)=A$ for any $A\in \X$.
However, the current definition of $\Gamma^*$ reveals that there is at least one explanation by selective salient limited attention of $c$, namely $(\Gamma^*,\rhd)$, such that $\Gamma^*(A)\subsetneq A$ holds for some $A\in \X$.}
To show that $\Gamma^*$ is a selective attention filter, toward a contradiction consider a menu $A\in\X$, and an item $x\in A$ such that $x\neq\max(\Gamma^*(A),\rhd)$ and $\Gamma^*(A\setminus x)\neq \Gamma^*(A)\setminus x $.
If $x=x_{\,\min}$, the definition of $\Gamma^*$ implies that $\Gamma^*(A)\setminus x=A\setminus{x}=\Gamma^*(A\setminus x)$, and we obtain a contradiction.
If $x\neq x_{\,\min}$,  the definition of $\Gamma^*$ yields $\Gamma^*(A)\setminus x=A\setminus \{x_{\,\min},x\}=\Gamma^*(A\setminus x)$, and we obtain a contradiction again.
Thus, condition (ii) of Definition~\ref{DEF:CSSLA} holds. 
To show that $(\Gamma^*,\rhd)$ is an explanation by selective salient limited attention of $c$, note that  $x_{\,\min}\neq c(A)=\max(A,\rhd)=\max(\Gamma^*(A),\rhd)$ for any $A\in\X\setminus \{x_{\,\min}\}$, and $c(x_{\,\min})=\max(\Gamma^*(x_{\,\min}),\rhd)=x_{\,\min}$.

\textit{\textbf{If part}}. Suppose $c\colon\X\to X$ is a choice with selective salient attention, and let $\left(\Gamma^{\,sl},\rhd\right)$ be some explanation by selective salient limited attention of $c$.
Assume toward a contradiction that Axiom\,$\alpha$ fails for $c$.
Thus, by Lemma \ref{LEMMA:minimal_violations_of_alpha} there are $A\in\X$ and $x\in  A$ such that $x\neq c(A)\neq c(A\setminus x)$. 
Definition~\ref{DEF:CSSLA} implies that $\Gamma^{\,s l}(A\setminus x)\neq\Gamma^{\,sl}(A)\setminus x$, which is impossible, since $x\neq c(A)=\max(\Gamma^{\,sl}(A),\rhd)$.

\medskip
\noindent{\underline{\it Proof of Theorem \ref{THM:list_rationalizable}.}
We first need some preliminary notions and results.
    
    \begin{definition}
Assume $c\colon\X\to X$ is a choice with competitive limited attention, and let $(\Gamma^{{\,co}},\rhd)$ be some explanation by competitive limited attention of $c$.
We say that $\Gamma^{\,co}$ is a \textsl{maximal competitive attention filter} for $c$ if there is no competitive attention filter $\Gamma^{{{\,co}}^{\prime}}\colon\X\to\X$ distinct from $\Gamma^{{\,co}}$ such that $(\Gamma^{{{\,co}}^{\prime}},\rhd)$ is an explanation by competitive limited attention of $c$, and $\Gamma^{{\,co}}(A)\subseteq \Gamma^{{\,co}^{\prime}}(A)$ for all $A \in \X$.\end{definition}

An analogue definition holds for choice with limited attention.

\begin{definition}
Assume $c\colon\X\to X$ is a choice with limited attention and $(\Gamma,\rhd)$ is some explanation by  limited attention of $c$.
We say that $\Gamma$ is a \textsl{maximal attention filter} for $c$ if there is no attention filter $\Gamma^{\prime}\colon\X\to\X$ distinct from $\Gamma$ such that $(\Gamma^{\prime},\rhd)$ is an explanation by limited attention of $c$, and $\Gamma(A)\subseteq \Gamma^{\prime}(A)$ for all $A \in \X$.\end{definition}

Thus, a maximal (competitive) attention filter describes the largest extent of DM's attention compatible with choice data.  
Let $\Gamma_\rhd (A) = \{x\in A\,\vert\, c(A)\rhd x\}\,\cup\,c(A)$. 
For choices with limited attention,  \citet[Lemma 17]{GiarlottaPetraliaWatson2022b} prove that $\Gamma_\rhd$ is a maximal attention filter.

\begin{lemma}\label{LEMMA:CLA_extension}
	Assume $c\colon \X\to X$ is a choice with limited attention, and let $P$ be the relation defined by $x P y$ if there is $A \in \X$ such that $x=c(A) \neq c(A \setminus y)$.
	 For any {strict} linear order $\rhd$ that extends $P$ we have that $\Gamma_\rhd$ is a maximal attention filter for $c$, and $(\Gamma_{\rhd},\rhd)$ is an explanation by  limited attention of $c$.
\end{lemma} 

A result analogue to Lemma 	\ref{LEMMA:CLA_extension} holds for choices with competitive limited attention.

\begin{lemma}\label{LEMMA:LR_maximal_filter}
	If $c$ is a choice with competitive limited attention , and $(\rhd, \Gamma^{\,co})$ is some explanation by competitive limited attention of $c$, then $\Gamma_{\rhd}$ is a maximal competitive attention filter for $c$, and $(\Gamma_{\rhd},\rhd)$ is an explanation by competitive limited attention of $c$.
\end{lemma}

\begin{proof}
	
We need to show that
	 (i) $c(A)=\max\left(\Gamma_{\rhd}(A),A\right)$ for any $A\in\X$, 
	(ii) $\Gamma_{\rhd}$ is a competitive attention filter, and
	 (iii) $\Gamma_{\rhd}$ is maximal for $c$.
	
\begin{enumerate}[\rm(i)]

\item The proof readily follows from the definition of $\Gamma_\rhd$.

\item We need to show that properties (a), and (b) listed in Definition \ref{DEF:list_attention} hold for $\Gamma_{\rhd}$.
\begin{itemize}
\item[(a)] Let $A \in \X$ be any menu, and $x\notin \Gamma_\rhd(A)$.
Toward a contradiction, suppose $\Gamma_\rhd(A)\setminus x \neq\Gamma_\rhd(A\setminus x)$.
The definition of $\Gamma_\rhd$ yields $\{y\in A\setminus x\,\vert\, c(A\setminus x)\rhd y\}\,\cup\,c(A\setminus x)\neq (\{y\in A\,\vert\, c(A)\rhd y\}\,\cup\,c(A))\setminus x,$ 
 hence $c(A)\neq c(A\setminus x)$.
Moreover, we have $x\neq \max(\Gamma(A),\rhd)$. 
Since $\Gamma^{\,co}(A)\setminus x=\Gamma^{\,co}(A \setminus x)$ because $\Gamma^{\,co}$ is a competitive attention filter, we obtain $c(A)\in \Gamma^{\,co}(A\setminus x)$ and $c(A\setminus x)\in\Gamma^{\,co}(A)$, which respectively yield $c(A\setminus x)\rhd c(A)$ and $c(A)\rhd c(A\setminus x)$, a contradiction. 
\item[(b)]($\Longrightarrow$). Let $y=\max(A,\rhd)$, $y \in \Gamma_\rhd(xy)$, and $x=\max(\Gamma_\rhd(A\setminus y),\rhd)$.
    Property (i) for $\Gamma_\rhd$ yields $c(xy)=y$ and $c(A\setminus y)=x$.
    Applying the same property to $\Gamma^{\,co}$, we get $y \in \Gamma^{\,co}(xy)$ and $x=\max(\Gamma^{\,co}(A\setminus y),\rhd)$.
    Since $\Gamma^{\,co}$ is a competitive attention filter, we conclude that $y \in \Gamma^{\,co}(A)$.
    Moreover, since $y=\max(A,\rhd)$ and $(\Gamma^{\,co},\rhd)$ is an explanation by competitive limited attention of $c$, we conclude that $y=c(A)$.
    The definition of $\Gamma_{\rhd}$ implies that $y \in \Gamma_\rhd(A)$.

($\Longleftarrow$). We prove this implication by contrapositive.
    Let $x=\max(\Gamma_\rhd(A\setminus x),\rhd)$, $y=\max(A,\rhd)$ and $y \notin \Gamma_\rhd(xy)$.
    Property (i) for $\Gamma_\rhd$ yields $c(xy)=x$ and $c(A\setminus y)=x$.
    Applying the same property to $\Gamma^{\,co}$, we get $y \notin \Gamma^{\,co}(xy)$ and $x=\max(\Gamma^{\,co}(A\setminus y),\rhd)$.
    Since $\Gamma^{\,co}$ is an explanation by competitive limited attention of $c$, we conclude that $y \notin \Gamma^{\,co}(A)$, and so $y \neq c(A)$.
    Toward a contradiction, suppose $y \in \Gamma_\rhd(A)$.
    Since $y=\max(A,\rhd)$, we conclude that $y=c(A)$, a contradiction.

\end{itemize}
\item Suppose by way of contradiction that there is a competitive attention filter ${\Gamma^{\,co}}^{\,\prime}\colon \X\to \X$ such that $({\Gamma^{\,co}}^{\,\prime},\rhd)$ is an  explanation by competitive limited attention of $c$ and $y\in {\Gamma^{\,co}}^{\,\prime}(A)\setminus \Gamma_\rhd(A)$ for some $A \in \X$ and $y\in A$.
Since $y\notin \Gamma_\rhd(A)$, we get $y\rhd c(A)$.
On the other hand, since $y \in{\Gamma^{\,co}}^{\,\prime}(A)$, $c$ is a choice with competitive limited attention, and $({\Gamma^{\,co}}^{\,\prime},\rhd)$ is an explanation by competitive limited attention of $c$, we must have $c(A)\rhd y$ or $c(A) = y$, which is impossible.
\end{enumerate}
\end{proof} 

The next lemma shows that maximal attention filters satisfy a specific regularity property.
\begin{lemma}\label{LEMMA:CLA_property}
		
	If $c$ is a choice with limited attention and $(\Gamma,\rhd)$ is some explanation by limited attention of $c$,	then for all $A \in \X$ we have $\Gamma_{\rhd}(A)=\Gamma_{\rhd}(\Gamma_{\rhd}(A))$.
\end{lemma}

\noindent \textit{\underline{Proof of Lemma \ref{LEMMA:CLA_property}.}}
Denote by $A_{*}$  the set
$$
A\setminus\{x_1,\cdots,x_i,\cdots, x_n\colon x_i \in (A\setminus\Gamma_\triangleright(A))\,(\forall\, 1\leq i\leq n)\}.
$$
Since $c$ is a choice with limited attention, and by Lemma \ref{LEMMA:CLA_extension} $\Gamma_{\rhd}$ is an attention filter, we have that $\Gamma_\triangleright (A)=\Gamma_\triangleright (A\setminus x_1)$.
Since $\Gamma_\triangleright (A)=\Gamma_\triangleright (A\setminus x_1)$, and $x_2\not \in \Gamma_\triangleright(A)$, we apply again the definition of attention filter to conclude that $\Gamma_\triangleright(A)=\Gamma_\triangleright (A\setminus x_1)=\Gamma_{\triangleright}(A\setminus x_1x_2)$.
We can repeat this argument until we get $\Gamma_{\triangleright}(A)=\cdots=\Gamma_{\triangleright}(A\setminus x_1\cdots x_{n})=\Gamma
_\triangleright(A\setminus A_{*})=\Gamma_{\triangleright}(\Gamma_\triangleright(A))$. 
\qed

Moreover, \citet[Corollary~1]{Yildiz2016} states that list rational choice functions are characterized by the asymmetry and acyclicity of $F_c$, introduced in Definition~\ref{DEF:revealed_list}, and a {strict} linear order that extends $F_c$ list rationalizes $c$.
The next result is taken from \citet[Corollary 1]{Yildiz2016}.

\begin{lemma}\label{LEMMA:LR_extension}
	If $c\colon\X\to X$ is list rational, and some list $\mathbf{f}$ list rationalizes $c$, then   $F_c \subseteq \mathbf{f}$.
\end{lemma}

The following result links choices with limited attention to list rationality.

\begin{lemma} \label{LEMMA: P_in_F}
	$P \subseteq F_c$.
\end{lemma}

\begin{proof}
	
	If $xPy$ holds for some $x,y\in X$, then there is $A\in\X$ such that $x=c(A)$ and $x \neq c(A \setminus y)$.
	Let $B=A\setminus y$ and observe that $x=c(B \cup y)$ while $x\neq c(A)$.
	Hence, $xF_cy$.
	\end{proof}

Finally we conclude:
    
\begin{corollary} \label{COR:LR_implies_CLA}
	If a choice is list rational, then it is a choice with limited attention.
\end{corollary}
 \begin{proof}
	Apply \citet[Corollary~1]{Yildiz2016} to obtain that $F_c$ is acyclic and antisymmetric.
	By Lemma~\ref{LEMMA: P_in_F}, $P$ is contained in $F_c$.
	Hence, $P$ is asymmetric and acyclic, and  \citet[Lemma~1 and Theorem~3]{MasatliogluNakajimaOzbay2012} implies that $c$ is a choice with limited attention.
	 \end{proof}

We are now ready to prove the equivalence between list rationality and choices with competitive limited attention.
\medskip
 
\textit{\textbf{Only if part}}.
Assume $c\colon\X\to X$ is list rational, and it is list rationalized by some list $\mathbf{f}$.    
    Thus, by Corollary~\ref{COR:LR_implies_CLA} $c$ is a choice with limited attention.
    Lemma~\ref{LEMMA:LR_extension} and Lemma~\ref{LEMMA: P_in_F} yield $P\subseteq \mathbf{f}$.
   By Lemma~\ref{LEMMA:CLA_extension} $\Gamma_{\mathbf{f}}$ is a maximal attention filter for $c$, and $\left(\Gamma_{\mathbf{f}},\mathbf{f}\right)$ is an explanation by limited attention of $c$. 
To show that $c$ is with competitive limited attention, we only need to prove that property (ii)(b) of Definition~\ref{DEF:list_attention} holds.

  \begin{description}
 \item ($\Longrightarrow$). Let $y=\max(A,\rhd)$, $y \in \Gamma_{\mathbf{f}}(xy)$, and $x=\max(\Gamma_{\mathbf{f}}(A\setminus y), \mathbf{f})$.
  	    Since $\left(\Gamma_{\mathbf{f}},\mathbf{f}\right)$ is an explanation by consideration of $c$, we have that $x=c(A\setminus y)$ and $y=c(xy)$.
  	    By hypothesis, $c$ is list rational and so $c(A)=c(c(A)y)=c(xy)=y$.
  	    Therefore $y \in \Gamma_{\mathbf{f}}(A )$.    
    
   \item($\Longleftarrow$). We again prove this implication by contrapositive.
       Let $y=\max(A,\rhd)$, $y \notin \Gamma_{\mathbf{f}}(xy)$, and $x=\max(\Gamma_{\mathbf{f}}(A\setminus y), \mathbf{f})$.
    Toward a contradiction, suppose $y \in \Gamma_\mathbf{f}(A)$.
     Since $\left(\Gamma_{\mathbf{f}},\mathbf{f}\right)$ is an explanation by consideration of $c$, we have that $x=c(A\setminus y)$ and $x=c(xy)$.
  	    By hypothesis, $c$ is list rational and so $c(A)=c(c(A)y)=c(xy)=x$.
  	    Therefore $x=\max(\Gamma_{\mathbf{f}}(A), \mathbf{f})$, a contradiction with $y \in \Gamma_\mathbf{f}(A)$ and $y \rhd x$.
\end{description}
    
    \textit{\textbf{If part}}.
    Suppose that $c\colon \X\to X$ is a choice with competitive limited attention, and  $(\Gamma^{\,co},\rhd)$ is some explanation by competitive limited attention of $c$.
    By Lemma~\ref{LEMMA:LR_maximal_filter}, $\Gamma_{\rhd}$ is a maximal competitive attention filter for $c$, and $(\Gamma_{\rhd},\rhd)$ is an explanation by competitive limited attention of $c$.
    We shall prove that $F_c \subseteq \triangleright$, and, by \citet[Corollary~1]{Yildiz2016}, $c$ is list rational, and $\rhd$ list rationalizes $c$.
         By contrapositive, suppose $\lnot(x \triangleright y)$ (thus $y \triangleright x$).
    Towards a contradiction, assume $xF_c y$.
    We prove that none of the three following properties, listed in Definition~\ref{DEF:revealed_list}, are true:
    \begin{itemize}
    	\item[\rm(i)] there is $A\in\X$ such that $x=c(A \cup y)$ and $y=c(xy)$;
    	\item[\rm(ii)] there is $A\in \X$ such that $x=c(A \cup y)$ and $x \neq c(A)$;
    	\item[\rm(iii)] there is $A\in \X$ such that $x\neq c(A\cup y)$, $x=c(xy)$, and $x=c(A)$.
    \end{itemize}
    
    \begin{enumerate}[\rm(i)]
    	\item 
    	Assume toward contradiction that there is $A\in\X$ such that $x=c(A \cup y)$ and $y=c(xy)$.
    	Note that by Lemma~\ref{LEMMA:CLA_property} we  can set without loss of generality $\Gamma_\triangleright(A \cup y)=\Gamma_\triangleright(A)=A$.\footnote{If $A\neq \Gamma_\triangleright(A)$, relabel $A$ as  $\{z\in A\,\vert\, x\rhd z\}\,\cup\,x$ and observe that now $\Gamma_\triangleright(A)=A$ and $x=c(A \cup y)$.}
   Moreover, $y \rhd \max(A,\rhd)$, $x=\max(\Gamma_\rhd(A \cup y),\rhd)=\max(\Gamma_\rhd(A),\rhd)=\max(A,\rhd)$, and $y \in \Gamma_\rhd(xy)$.
   Since $\Gamma_{\rhd}$ is a competitive attention filter, by condition (ii)(b) of Definition~\ref{DEF:list_attention} we obtain $y \in \Gamma_\rhd(A \cup y)$, a contradiction with $x=c(A \cup y)$.
        
\item Assume toward a contradiction that there are $A\in\X$, $y\in X$, and $x\in A$ such that $x=c(A	\cup y)$ and $x\neq c(A)$.
    Since $\Gamma_{\rhd}$ is a competitive attention filter, and $y\not\in \Gamma_{\triangleright}(A\cup y)$, by condition (ii)(a) of Definition~\ref{DEF:list_attention} we have $\Gamma_{\triangleright}(A\cup y)=\Gamma_{\triangleright}(A)$, that yields $c(A\cup y)=c(A)$, which is false.
    
    \item Assume toward a contradiction that there are $A\in\X$, $y\in X$, and $x\in A$ such that $c(A\cup y)\neq x$, $c(A)=x$, and $c(xy)=x$.
        Choose an $A$ of minimal cardinality such that $c(A)=x$ and $x \neq c(A \cup y)$.
        
        If $y \notin \Gamma_\triangleright(A \cup y)$, since $\Gamma_{\rhd}$ is a competitive attention filter, by property (ii)(a) of Definition~\ref{DEF:list_attention} we have that $\Gamma_\triangleright(A \cup y)=\Gamma_\triangleright(A)$.
     By Definition~\ref{DEF:list_attention} again we obtain  $x=c(A\cup y)$,  
         which is false. 
     
     If $y\in\Gamma_\triangleright(A \cup y)$ and $y \rhd \max(A,\rhd)$, then by property (ii)(b) of Definition~\ref{DEF:list_attention} we must have $y \in \Gamma_\rhd(xy)$,  which yields $c(xy)=y$, a contradiction.
     
     Assume now that $y\in\Gamma_\triangleright(A \cup y)$ and  there is a $z \in A$ such that $\max(A,\rhd)=z \rhd y$.
     Note that $z \neq x$, because $y \rhd x$.
     Since $x=c(A)$,  we know that $z \notin \Gamma_\rhd(A)$, which implies that that $\Gamma_\rhd(A)=\Gamma_\rhd(A\setminus z)$ and $x=c(A\setminus z)$.
     Summarizing what we found, $z\rhd\max(A \setminus z,\rhd)$, $x=c(A\setminus z)$ and $z \notin \Gamma_\rhd(((A\setminus z) \cup z))$.
     By condition (ii)(b) of Definition~\ref{DEF:list_attention}, we have that $z \notin \Gamma_\rhd(xz)$.
     Two cases are possible: $x=c(((A \cup y)\setminus z))$, or $x\neq c(((A\setminus z) \cup y))$.
     If $x=c(((A \cup y)\setminus z))$, then $x=\max(\Gamma_{\rhd}(((A \cup y)\setminus z)),\rhd)$.
     Since $z \notin \Gamma_\rhd(xz)$, we can apply again condition (ii)(b) of Definition~\ref{DEF:list_attention} to obtain $z \notin \Gamma_\rhd(((A \cup y)\setminus z) \cup z)=\Gamma_\rhd(A \cup y)$.
     Therefore $\Gamma_\rhd(A \cup y)= \Gamma_\rhd((A \cup y)\setminus z)$, which contradicts the fact that $x \neq c(A \cup y)$.
     We conclude that $x\neq c(((A\setminus z) \cup y))$ and $x=c(A\setminus z)$, which contradicts minimality of $A$.
     
     \end{enumerate}
     
    Therefore, none of the three properties hold, and $\neg (xF_c y)$, which contradicts our hypothesis.
    We conclude that $\lnot(x \triangleright y)$ implies $\lnot(xF_c y)$.
    Hence, $F_c \subseteq \triangleright$, which implies that $F_c$ is asymmetric and acyclic.
    By \citet[Corollary~1]{Yildiz2016}, $c$ is list rational.\footnote{Note that Lemma \ref{LEMMA:LR_extension} implies that $\rhd$ list rationalizes $c$.}
    \qed

\medskip
\noindent\underline{\it Proof of Lemma~\ref{LEMMA:necessary_conditions_of_CCLA}.}
By Theorem~\ref{THM:list_rationalizable} $c$ is also list rational.
Observe that if $x\neq c(A)\neq c(A\setminus x)$, by case  (ii) of Definition~\ref{DEF:revealed_list} we have that $c(A)F_c x$.
Since \citet[Corollary~1]{Yildiz2016} shows that any list $\mathbf{f}$ that list rationalizes a choice function $c$ is a transitive closure of $F_c$, we conclude that $c(A)\,\mathbf{f}x$.
 Moreover, Definition~\ref{DEF:list_attention} implies that $\Gamma^{\,co}(A)\neq \Gamma^{\,co}(A\setminus x)$.
Condition (ii)(a) of Definition~\ref{DEF:list_attention} yields $x\in\Gamma^{\,co}(A)$.
Since $x\in\Gamma^{\,co}(A)$, we conclude that $c(A)\rhd x$.

Assume now that there are $x,y\in X$, and $A\in\X$ such that $x=c(xy)=c(A\setminus y) \neq c(A)=y$.
By case  (iii) of Definition~\ref{DEF:revealed_list} we have that $xF_cy$, and, by applying again \citet[Corollary~1]{Yildiz2016}, that $x\,\mathbf{f}y$.
Moreover, assume toward a contradiction that there is an explanation by competitive limited attention $(\Gamma^{\,co},\rhd)$ of $c$ such that $y=\max(A,\rhd)$.
By property (ii)(b) of Definition~\ref{DEF:list_attention} we obtain that  $y\in\Gamma^{\,co}(xy)$, and that $x\rhd y$, a contradiction.

Finally, suppose that there are $x,y\in X$, and $A\in\X$ such that $y=c(xy)\neq c(A\setminus y)=c(A)=x$.
By case  (i) of Definition~\ref{DEF:revealed_list} we have that $xF_c y$, and, by applying again \citet[Corollary~1]{Yildiz2016}, that $x\,\mathbf{f}y$.
Moreover, assume toward a contradiction that there is an explanation by competitive limited attention $(\Gamma^{\,co},\rhd)$ of $c$ such that $y=\max(A,\rhd)$ holds.
By property (ii)(b) of Definition~\ref{DEF:list_attention} we obtain that  $y\in\Gamma^{\,co}(A)$, and that $x\rhd y$, a contradiction. \qed


\begin{thebibliography}{31}

\providecommand{\natexlab}[1]{#1}
\providecommand{\url}[1]{\texttt{#1}}
\expandafter\ifx\csname urlstyle\endcsname\relax 
  \providecommand{\doi}[1]{doi: #1}\else
  \providecommand{\doi}{doi: \begingroup \urlstyle{rm}\Url}\fi




\bibitem[Aguiar et al.(2023)]{Aguiaretal2023}{\textsc{Aguiar, V.\,H., Boccardi, M.\,J., Kashaev, N., and Kim, J.}, 2023.
Random utility and limited consideration.
\textit{Quantitative Economics}, 14:\,71--116.}



\bibitem[Allenby and Ginter(1995)]{AllenbyGinter1995}{\textsc{Allenby, G., and Ginter, J.}, 1995.
The effects of in-store displays and feature advertising on consideration sets.
\textit{Journal of Research in Marketing}, 12:\,67--80.}


\bibitem[Bordalo, Gennaioli, and Shleifer(2012)]{BordaloGennaioliShleifer2012}{\textsc{Bordalo, P., Gennaioli, N., and Shleifer, A.}, 2012.
Salience theory of choice under risk. 
\textit{Quarterly Journal of Economics} 127:\,1243--1285.}

\bibitem[Bordalo, Gennaioli, and Shleifer(2013)]{BordaloGennaioliShleifer2013}{\textsc{Bordalo, P., Gennaioli, N., and Shleifer, A.}, 2013.
Salience and consumer choice. 
\textit{Journal of Political Economy}, 121:\,803--843.}




\bibitem[Cantone et al.(2016)]{CantoneGiarlottaGrecoWatson2016}{\textsc{Cantone, D., Giarlotta, A., Greco, S., and Watson, S.}, 2016.
$(m,n)$-rationalizable choices.
\textit{Journal of Mathematical Psychology},
73:\,12--27.}

\bibitem[Caplin and Dean(2015)]{CaplinDean2015}{\textsc{Caplin, A., and Dean, M.}, 2015.
Revealed preference, rational inattention, and costly information acquisition.
\textit{American Economic Review},
105:\,2183--2203.}



\bibitem[Caplin, Dean, and Leahy(2019)]{CaplinDeanLeahy2019}{\textsc{Caplin, A., Dean, M., and Leahy, J.}, 2019.
Rational inattention, optimal consideration sets, and stochastic choice.
\textit{The Review of Economic Studies}, 86:\,1061--1094.}


\bibitem[Cattaneo et al.(2020)]{Cattaneoetal2020}{\textsc{Cattaneo, M.\, D., Ma, X., Masatlioglu, Y., and Suleymanov, E.}, 2020.
A random attention model.
\textit{Journal of Political Economy}, 128:\,2796--2836.
}


\bibitem[Cherepanov, Feddersen, and Sandroni(2013)]{CherepanovFeddersenSandroni2013}
{\textsc{Cherepanov, V., Feddersen, T., and Sandroni, A.}, 2013.
Rationalization. 
\textit{Theoretical Economics}, 8:\,775--800.}

\bibitem[Chernoff(1954)]{Chernoff1954}
{\textsc{Chernoff, H.}, 1954. 
Rational selection of decision functions. 
\textit{Econometrica}, 22:\,422--443.}


\bibitem[Chiang, Ching, and Narasimhan(1999)]{ChiangChibNarasimhan1999}{\textsc{Chiang, J., Chib, S., and Narasimhan, C.}, 1999.
Markov chain Monte Carlo and models of consideration set and parameter heterogeneity.
\textit{Journal of Econometrics}, 89:\,223--248.}





\bibitem[Dutta and Horan(2015)]{DuttaHoran2015}{\textsc{Dutta, R., and Horan, S.}, 2015. Inferring Rationales from Choice: Identification for rational shortlist methods.
\textit{American Economic Journal: Microeconomics}, 7:\,179--201.}



\bibitem[Fader and McAlister(1990)]{FaderMcAlister1990}{\textsc{Fader, P., and McAlister, L.}, 1990.
An elimination by aspects model of consumer response to promotion calibrated on UPC scanner data.
\textit{Journal of Marketing Research}, 27:\,322--332.}


\bibitem[Giarlotta, Petralia, and Watson(2022)]{GiarlottaPetraliaWatson2022a}{\textsc{Giarlotta, A., Petralia, A.\,E., and Watson, S.}, 2022.
Bounded rationality is rare.
\textit{Journal of Economic Theory}, 204, 105509.} 

\bibitem[Giarlotta, Petralia, and Watson(2023)]{GiarlottaPetraliaWatson2022b}
{\textsc{Giarlotta, A., Petralia, A.\,E., and Watson, S.}, 2023.
Context-Sensitive Rationality: Choice by Salience.
\textit{Journal of Mathematical Economics}, 109, 102913.}


\bibitem[Gilbride and Allenby(2004)]{GilbrideAllenby2004}{\textsc{Gilbride, T., and Allenby, G.}, 2004.
A choice model with conjunctive, disjunctive, and compensatory screening rules.
\textit{Marketing Science}, 23:\,391--406.}





\bibitem[Kibris, Masatlioglu, and Suleymanov(2021)]{KibrisMasatliogluSuleymanov2021}
{\textsc{Kibris, O., Masatlioglu, Y., and Suleymanov, E.}, 2021.
A Theory of Reference Point Formation.
\textit{Economic Theory}.}



\bibitem[Lanzani(2022)]{Lanzani2022}{\textsc{Lanzani, G.}, 2022.
Correlation made simple: application to salience and regret theory.
\textit{The Quaterly Journal of Economics}, 137:\,959--987.}




\bibitem[Lleras, Masatlioglu, Nakajima, and Ozbay(2017)]{LlerasMasatliogluNakajimaOzbay2017}
{\textsc{Lleras, J.,\, S, Masatlioglu, Y., Nakajima, D., and Ozbay, E.\,Y.}, 2017.
When more is less: limited consideration.
\textit{Journal of Economic Theory}, 170:\,70--85.}

\bibitem[Lleras, Masatlioglu, Nakajima, and Ozbay(2021)]{LlerasMasatliogluNakajimaOzbay2021}
{\textsc{Lleras, J.,\, S, Masatlioglu, Y., Nakajima, D., and Ozbay, E.\,Y.}, 2021.
Path-Independent consideration.
\textit{Games}, 12,\,21.} 



\bibitem [Manzini and Mariotti(2007)]{ManziniMariotti2007} 
{\textsc{Manzini, P., and Mariotti, M.}, 2007.
Sequentially rationalizable choice. 
\textit{American Economic Review}, 97:\,1824--1839.}



\bibitem[Manzini and Mariotti(2014a)]{ManziniMariotti2014}{\textsc{Manzini, P., and Mariotti, M.}, 2014.
Stochastic choice and consideration sets.
\textit{Econometrica}, 82:\,1153-1176.}

\bibitem[Manzini and Mariotti(2014b)]{ManziniMariotti2014b}{\textsc{Manzini, P., and Mariotti, M.}, 2014.
Welfare economics and bounded rationality: the case
for model-based approaches.
\textit{Journal of Economic Methodology}, 21:\,343-360.}


%
\bibitem[Masatlioglu, Nakajima, and Ozbay(2012)]{MasatliogluNakajimaOzbay2012}
{\textsc{Masatlioglu, Y., Nakajima, D., and Ozbay, E.\,Y.}, 2012.
Revealed attention. 
\textit{American Economic Review}, 102:\,2183--2205.}



%

\bibitem[Parr and Friston(2017)]{ParrandFriston2017}{\textsc{Parr, T., and Friston, K.\,J.} 2017.
Working memory, attention, and salience in active inference.
\textit{Scientific Reports}, 7:\,14678.}

\bibitem[Parr and Friston(2019)]{ParrandFriston2019}{\textsc{Parr, T., and Friston, K.\,J.} 2019.
Attention or Salience?
\textit{Current Opinion in Psychology}, 29:\,1--5.}


\bibitem[Paulssen and Bagozzi(2005)]{PaulssenBagozzi2005}{\textsc{Paulssen, M, and Bagozzi, R.\,P.}, 2005.
A self-regulatory model of consideration set formation.
\textit{Psychology and \& Marketing}, 22:\,785--812.}




\bibitem[Ravid and Steverson(2021)]{RavidStevenson2021}{\textsc{Ravid, D., and Steverson, K.}, 2021.
Bad temptation.
\textit{Journal of Mathematical Economics}, 95:\,102480.}


\bibitem[Roberts and Lattin(1991)]{RobertsLattin1991}{\textsc{Roberts, J., and Lattin, J.}, 1991.
Development and testing of a model of consideration set composition.
\textit{Journal of Marketing Research}, 28:\,429--440.}


\bibitem[Samuelson(1938)]{Samuelson1938}
{\textsc{Samuelson, A.\,P.}, 1938.
A note on the pure theory of consumer's behaviour.
\textit{Economica}, 17:\,61--71.} 
%


\bibitem[Sen(1971)]{Sen1971}{\textsc{Sen, A.\,K.}, 1971.
Choice functions and revealed preference. 
\textit{The Review of Economic Studies}, 38:\,307--317.}

\bibitem[Sen(1986)]{Sen1986}{\textsc{Sen, A.\,K.}, 1986.
Social choice theory.
In K.\,J. Arrow and \& M.\,D. Intrilligator (Eds.),
\textit{Handbook of Mathematical Economics Vol III} (pp. 1073--1081). Elsevier Science Publisher, North-Holland.}



\bibitem[Smith and Krajbich(2018)]{SmithKrajbic2018}{\textsc{Smith, S.\,M., and Krajbich, I.}, 2018.
Attention and choice across domains.
\textit{Journal of Experimental Psychology: General}, 147:\, 1810--1826.}




\bibitem[Suh(2009)]{Suh2009}{\textsc{Suh, J.\,C.}, 2009.
The role of consideration sets in brand choice: The moderating role of product characteristics.
\textit{Psychology \& Marketing}, 26:\,534--550.}


\bibitem[Szpilrajn(1930)]{Szpilrajn1930}{\textsc{Szpilrajn, E.}, 1930.
Sur l'extension de l'ordre partiel. 
\textit{Fundamenta Mathematicae}, 16: 386--389.}

\bibitem[Terui, Ban, and Allenby(2011)]{TeruiBanAllenby2021}{\textsc{Terui, N., Ban, M., and Allenby, G.\,M.}, 2011.
Advertising on brand consideration and choice.
\textit{Marketing Science}, 30:\,74--91.}


\bibitem[Tyson(2013)]{Tyson2013}{\textsc{Tyson, C.\,J.}, 2013.
Behavioral implications of shortlisting procedures.
\textit{Social Choice and Welfare}, 41:\,941--963.}


\bibitem[Van Nierop et al.(2010)]{VanNieropetal}{\textsc{Van Nierop, E., Bronnenberg, B., Paap, R., Wedel, M., and Franses, P.\,H.}, 2010.
Retrieving unobserved consideration sets from household panel data.
\textit{Journal of Marketing Research}, 47:\,63--74.}


\bibitem[Yildiz(2016)]{Yildiz2016}
{\textsc{Yildiz, K.}, 2016.
List-rationalizable choice.
\textit{Theoretical Economics}, 11:\,587--599.}




\end{thebibliography}
 \end{document}